\documentclass[twocolumn,tighten,astrosymb,trackchanges]{aastex631}
\newcommand{\LCO}{\affiliation{Las Cumbres Observatory, 6740 Cortona Drive, Suite 102, Goleta, CA 93117-5575, USA}}
\newcommand{\UCSB}{\affiliation{Department of Physics, University of California, Santa Barbara, CA 93106-9530, USA}}

\newcommand{\UCD}{\affiliation{Department of Physics and Astronomy, University of California, Davis, 1 Shields Avenue, Davis, CA 95616-5270, USA}}
\newcommand{\WIS}{\affiliation{Department of Particle Physics and Astrophysics, Weizmann Institute of Science, 76100 Rehovot, Israel}}
\newcommand{\OKC}{\affiliation{Department of Physics, Oskar Klein Centre, Stockholm University, SE-106 91, Stockholm, Sweden}}
\newcommand{\OKCAstro}{\affiliation{Department of Astronomy, Oskar Klein Center, Stockholm University, SE-106 91 Stockholm, Sweden}}

\newcommand{\CaltechPhys}{\affiliation{Division of Physics, Mathematics and Astronomy, California Institute of Technology, Pasadena, CA 91125, USA}}

\newcommand{\UCB}{\affiliation{Department of Astronomy, University of California, Berkeley, CA 94720-3411, USA}}

\newcommand{\STScI}{\affiliation{Space Telescope Science Institute, 3700 San Martin Drive, Baltimore, MD 21218-2410, USA}}
\newcommand{\UT}{\affiliation{Department of Astronomy, The University of Texas at Austin, 2515 Speedway, Stop C1400, Austin, TX 78712, USA}}

\newcommand{\Tsinghua}{\affiliation{Physics Department and Tsinghua Center for Astrophysics, Tsinghua University, Beijing, 100084, People's Republic of China}}

\newcommand{\CfA}{\affiliation{Center for Astrophysics \textbar{} Harvard \& Smithsonian, 60 Garden Street, Cambridge, MA 02138-1516, USA}}
\newcommand{\IAIFI}{\affiliation{The NSF AI Institute for Artificial Intelligence and Fundamental Interactions, USA}}
\newcommand{\UA}{\affiliation{Steward Observatory, University of Arizona, 933 North Cherry Avenue, Tucson, AZ 85721-0065, USA}}

\newcommand{\Carnegie}{\affiliation{Observatories of the Carnegie Institute for Science, 813 Santa Barbara Street, Pasadena, CA 91101-1232, USA}}

\newcommand{\UCSC}{\affiliation{Department of Astronomy and Astrophysics, University of California, Santa Cruz, CA 95064-1077, USA}}
\newcommand{\Purdue}{\affiliation{Department of Physics and Astronomy, Purdue University, 525 Northwestern Avenue, West Lafayette, IN 47907-2036, USA}}
\newcommand{\Princeton}{\affiliation{Department of Astrophysical Sciences, Princeton University, 4 Ivy Lane, Princeton, NJ 08540-7219, USA}}

\newcommand{\JHU}{\affiliation{Department of Physics and Astronomy, The Johns Hopkins University, 3400 North Charles Street, Baltimore, MD 21218, USA}}

\newcommand{\CIERA}{\affiliation{Center for Interdisciplinary Exploration and Research in Astrophysics (CIERA), 1800 Sherman Ave., Evanston, IL 60201, USA}}
\newcommand{\Northwestern}{\affiliation{Department of Physics and Astronomy, Northwestern University, 2145 Sheridan Rd, Evanston, IL 60208, USA}}
\newcommand{\SkAI}{\affiliation{NSF-Simons AI Institute for the Sky (SkAI), 172 E. Chestnut St., Chicago, IL 60611, USA}}

\newcommand{\IIA}{\affiliation{Indian Institute of Astrophysics, Koramangala 2nd Block, Bangalore 560034, India}}

\newcommand{\GeminiNorth}{\affiliation{Gemini Observatory/NSF's NOIRLab, 670 North A`ohoku Place, Hilo, HI 96720-2700, USA}}

\newcommand{\Rutgers}{\affiliation{Department of Physics and Astronomy, Rutgers, the State University of New Jersey,\\136 Frelinghuysen Road, Piscataway, NJ 08854-8019, USA}}

\newcommand{\Melbourne}{\affiliation{School of Physics, The University of Melbourne, Parkville, VIC 3010, Australia}}

\newcommand{\ESOgermany}{\affiliation{European Southern Observatory, Karl-Schwarzschild-Stra\ss{}e 2, D-85748, Garching bei M\"unchen, Germany}}
\newcommand{\ICE}{\affiliation{Institute of Space Sciences (ICE, CSIC), Campus UAB, Carrer de Can Magrans, s/n, E-08193 Barcelona, Spain}}
\newcommand{\IEEC}{\affiliation{Institut d’Estudis Espacials de Catalunya (IEEC), E-08034 Barcelona, Spain}}

\newcommand{\USzeged}{\affiliation{Department of Experimental Physics, Institute of Physics, University of Szeged, D\'om t\'er 9, 6720 Szeged, Hungary}}
\newcommand{\BAOUSzeged}{\affiliation{Baja Astronomical Observatory of the University of Szeged, Szegedi {\'u}t, Kt. 766, 6500 Baja, Hungary}}

\newcommand{\MTAELTE}{\affiliation{MTA-ELTE Lend\"ulet "Momentum" Milky Way Research Group, Hungary}}
\newcommand{\Konkoly}{\affiliation{Konkoly Observatory, Research Centre for Astronomy and Earth Sciences (CSFK), MTA Center of Excellence, Konkoly-Thege Mikl\'os \'ut 15-17, 1121 Budapest, Hungary}}

\newcommand{\AMNH}{\affiliation{Department of Astrophysics, American Museum of Natural History, Central Park West and 79th Street, New York, NY 10024-5192, USA}}
\newcommand{\UDublin}{\affiliation{School of Physics, Trinity College Dublin, The University of Dublin, Dublin 2, Ireland}}

\newcommand{\ICG}{\affiliation{Institute of Cosmology and Gravitation, University of Portsmouth, Dennis Sciama Building, Burnaby Road, Portsmouth PO1 3FX, UK}}

\newcommand{\GSI}{\affiliation{GSI Helmholtzzentrum f\"ur Schwerionenforschung, Planckstra\ss{}e 1, 64291 Darmstadt, Germany}}
\newcommand{\KyotoU}{\affiliation{Department of Astronomy, Kyoto University, Kitashirakawa-Oiwake-cho, Sakyo-ku, Kyoto, 606-8502. Japan}}

\newcommand{\NotreDame}{\affiliation{Department of Physics and Astronomy, University of Notre Dame, Notre Dame, IN 46556, USA}}

\newcommand{\AixMarseille}{\affiliation{Aix Marseille Univ, CNRS, CNES, LAM, Marseille, France}}

\newcommand{\Thailand}{\affiliation{National Astronomical Research Institute of Thailand, 260 Moo 4, Donkaew, Maerim, Chiang Mai 50180, Thailand}}
\newcommand{\UAB}{\affiliation{Instituto de Astrofísica, Universidad Andres Bello, Fernandez Concha 700, Las Condes, Santiago RM, Chile}}
\newcommand{\PUC}{\affiliation{Instituto de Astrof\'{i}sica, Facultad de F\'{i}sica, Pontificia Universidad Cat\'{o}lica de Chile, Av. Vicu\~{n}a Mackenna 4860, Santiago, Chile}}
\newcommand{\MIA}{\affiliation{Millennium Institute of Astrophysics, Nuncio Monse\~{n}or S\'{o}tero Sanz 100, Providencia, Santiago, Chile}}

\newcommand{\ELTE}{\affiliation{ELTE E\"otv\"os Lor\'and University, Institute of Physics and Astronomy, P\'azm\'any P\'eter s\'et\'any 1/A, Budapest, 1117 Hungary}}

\newcommand{\UVa}{\affiliation{Department of Astronomy, University of Virginia, 530 McCormick Rd, Charlottesville, VA 22904, USA}}
\newcommand{\CaltechOO}{\affiliation{Caltech Optical Observatories, California Institute of Technology, Pasadena, CA 91125, USA}}
\newcommand{\NOT}{\affiliation{Nordic Optical Telescope, Rambla José Ana Fernández Pérez 7, ES-38711 Breña Baja, Spain}}
\newcommand{\ARIES}{\affiliation{Aryabhatta Research Institute of Observational Sciences (ARIES), Manora Peak, Nainital - 263001, India}}
\newcommand{\ANU}{\affiliation{Research School of Astronomy and Astrophysics, Australian National University, Canberra, ACT 0200, Australia}}

\newcommand{\UCO}{\affiliation{University of California Observatories, 550 Red Hill Rd, Santa Cruz, CA 95064, USA}}
\newcommand{\Lick}{\affiliation{UCO/Lick Observatory, PO Box 85, Mount Hamilton, CA 95140, USA}}
\newcommand{\ZhejiangU}{\affiliation{Institute for Advanced Study in Physics, Zhejiang University, Hangzhou 310027, China}}

\newcommand{\UIUC}{\affiliation{Department of Astronomy, University of Illinois, Urbana, IL 61801, USA}}
\newcommand{\NCSA}{\affiliation{Center for Astrophysical Surveys, National Center for Supercomputing Applications, Urbana, IL 61801, USA}}
\newcommand{\ICASU}{\affiliation{Illinois Center for Advanced Studies of the Universe; Urbana, IL 61801, USA}}
\newcommand{\UTurku}{\affiliation{Tuorla Observatory, Department of Physics and Astronomy, FI-20014 University of Turku, Finland}}
\newcommand{\FINCA}{\affiliation{Finnish Centre for Astronomy with ESO (FINCA), FI-20014 University of Turku, Finland}}
\newcommand{\CITEVA}{\affiliation{Centro de Astronomía (CITEVA), Universidad de Antofagasta, Av. Angamos 601, Antofagasta, Chile}}

\newcommand{\UChicago}{\affiliation{Department of Astronomy and Astrophysics, University of Chicago, William Eckhart Research Center, 5640 South Ellis Avenue, Chicago, IL 60637, USA}}
\usepackage{CJK}
\usepackage{multirow}
\newcommand\kms{~km~s$^{-1}$}

\newcommand\um{~$\mu$m}
\let\tablenum\relax
\usepackage{siunitx}
\usepackage{appendix}
\usepackage{amsmath,amsthm,amsfonts,amssymb,amscd}
\usepackage{booktabs}
\usepackage{tablefootnote}
\usepackage{apjfonts}
\usepackage{lipsum}
\usepackage{changepage}

\let\ts=\thinspace
\newcommand{\one}{\ts {\sc i}}
\newcommand{\two}{\ts {\sc ii}}
\newcommand{\three}{\ts {\sc iii}}
\newcommand{\four}{\ts {\sc iv}}
\newcommand{\five}{\ts {\sc v}}
\newcommand{\six}{\ts {\sc vi}}
\newcommand{\seven}{\ts {\sc vii}}
\renewcommand\ion[2]{#1\,\,{\sc{\romannumeral #2}}}

\definecolor{maroon}{rgb}{0.760,0.118,0.337}

\definecolor{darkaqua}{rgb}{0.0,0.45,0.65}

\def\msun{\hbox{\,$M_{\odot}$}}

\def\cm{\mbox{\,cm}}
\def\cm3{\mbox{\,cm$^{-3}$}}

\shorttitle{JWST SN~Iax}
\shortauthors{Kwok et al.}
\begin{document}

\title{JWST and Ground-based Observations of the Type Iax Supernovae SN~2024pxl and SN~2024vjm: \\ Evidence for Weak Deflagration Explosions}

\correspondingauthor{Lindsey A. Kwok}
\email{lindsey.kwok@northwestern.edu}

% %% first 3 will get shuffled for papers I, II, III
\author[0000-0003-3108-1328]{Lindsey A.\ Kwok}
\thanks{CIERA Fellow}
\CIERA

\author[0000-0001-6706-2749]{Mridweeka Singh}
\IIA

%% all the rest should be consistent across all three papers
\author[0000-0001-8738-6011]{Saurabh W.\ Jha}
\Rutgers

\author[0000-0002-9388-2932]{St\'{e}phane Blondin}
\AixMarseille, \ESOgermany

\author[0000-0001-6191-7160]{Raya~Dastidar}
\UAB
\MIA

\author[0000-0003-2037-4619]{Conor~Larison}
\Rutgers

\author[0000-0001-9515-478X]{Adam~A.~Miller}
\Northwestern
\CIERA
\SkAI

\author[0000-0003-0123-0062]{Jennifer E.\ Andrews}
\GeminiNorth

\author[0000-0002-1895-6639]{Moira~Andrews}
\LCO
\UCSB

\author[0000-0003-3533-7183]{G. C. Anupama}
\IIA

% \author[0000-0002-6688-3307]{Prasiddha Arunachalam}
% \UCSC

\author[0000-0002-4449-9152]{Katie Auchettl}
\Melbourne
\UCSC

\author[0000-0001-9275-0287]{Dominik B\'anhidi}
\USzeged
\BAOUSzeged

\author[0000-0003-4769-4794]{Barnabas Barna}
\USzeged

\author[0000-0002-4924-444X]{K.\ Azalee Bostroem}
\thanks{LSST-DA Catalyst Fellow}
\UA

\author[0000-0001-5955-2502]{Thomas G.\ Brink}
\UCB

\author[0000-0003-4553-4033]{R\'egis Cartier}
\CITEVA

\author[0000-0003-0853-6427]{Ping Chen}
\ZhejiangU
\WIS

\author[0000-0003-0528-202X]{Collin~T.~Christy}
\UA

\author[0000-0003-4263-2228]{David~A.~Coulter}
\STScI

\author[0000-0003-1858-561X]{Sofia~Covarrubias}
\CaltechPhys

\author[0000-0002-5680-4660]{Kyle W.\ Davis}
\UCSC

\author[0000-0001-9749-4200]{Connor~B.~Dickinson}
\UCSC

\author[0000-0002-7937-6371]{Yize Dong}
\CfA

\author[0000-0003-4914-5625]{Joseph~R.~Farah}
\LCO
\UCSB

\author[0000-0003-3460-0103]{Alexei V.\ Filippenko}
\UCB
%\Hagler

\author[0000-0003-2024-2819]{Andreas Fl\"ors}
\GSI

\author[0000-0002-2445-5275]{Ryan J.\ Foley}
\UCSC

\author[0000-0003-4537-3575]{Noah Franz}
\UA

\author[0000-0002-4223-103X]{Christoffer~Fremling}
\CaltechPhys
\CaltechOO

\author[0000-0002-1296-6887]{Llu\'is Galbany}
\ICE
\IEEC

\author[0000-0002-3884-5637]{Anjasha Gangopadhyay}
\OKCAstro

\author[0009-0002-4441-3192]{Aarna~Garg}
\UCSC

\author[0000-0003-4069-2817]{Peter Garnavich}
\NotreDame

\author[0000-0002-3739-0423]{Elinor~L.~Gates}
\Lick

\author[0000-0002-4391-6137]{Or~Graur}
\ICG
\AMNH

\author[0000-0002-5025-4645]{Alexa~C.~Gordon}
\Northwestern
\CIERA

\author[0000-0002-1125-9187]{Daichi~Hiramatsu}
\CfA
\IAIFI

\author[0000-0003-2744-4755]{Emily~Hoang}
\UCD

% \author[0000-0002-0832-2974]{Griffin Hosseinzadeh}
% \UCSD

\author[0000-0003-4253-656X]{D.~Andrew~Howell}
\LCO
\UCSB

\author[0000-0002-9454-1742]{Brian Hsu}
\UA

\author[0000-0001-5975-290X]{Joel Johansson}
\OKC

\author[0000-0001-9275-0287]{Arti Joshi}
\PUC

\author[0009-0007-5296-4046]{Lordrick~A.~Kahinga}
\UCSC

\author[0009-0005-1871-7856]{Ravjit Kaur}
\UCSC

\author[0000-0001-8367-7591]{Sahana~Kumar}
\UVa

\author[0009-0004-7572-5679]{Piramon~Kumnurdmanee}
\UCSC

\author[0000-0002-1132-1366]{Hanindyo Kuncarayakti}
\UTurku
\FINCA

\author[0000-0002-2249-0595]{Natalie~LeBaron}
\UCB

% \author[0000-0003-1731-0497]{C.~Lidman}
% \ANU
% \ANUCGA

\author[0000-0002-7866-4531]{Chang~Liu}
\Northwestern
\CIERA

\author[0000-0003-2611-7269]{Keiichi Maeda}
\KyotoU

\author[0000-0002-9770-3508]{Kate Maguire}
\UDublin

% \author[0009-0006-4963-3206]{Bailey Martin}
% \ANU

\author[0000-0001-5807-7893]{Curtis McCully}
\LCO
\UCSB

\author[0009-0008-9693-4348]{Darshana Mehta}
\UCD

\author[0000-0001-7771-4624]{Luca~M.~Menotti}
\UCSC

\author[0009-0007-8154-6863]{Anne~J.~Metevier}
\UCO

\author[0000-0003-1637-267X]{Kuntal Misra}
\ARIES

\author[0009-0006-5214-0736]{C.~Tanner~Murphey}
\UIUC
\NCSA
\ICASU

\author[0000-0001-9570-0584]{Megan Newsome}
\LCO
\UT

\author[0000-0003-0209-9246]{Estefania Padilla~Gonzalez}
\JHU

\author[0000-0002-1092-6806]{Kishore~C.~Patra}
\UCSC

\author[0000-0002-0744-0047]{Jeniveve Pearson}
\UA

\author[0000-0001-6806-0673]{Anthony~L.~Piro}
\Carnegie

\author[0000-0002-1633-6495]{Abigail~Polin}
\Purdue

\author[0000-0002-7352-7845]{Aravind~P.~Ravi}
\UCD

\author[0000-0002-4410-5387]{Armin Rest}
\STScI

\author[0000-0002-5683-2389]{Nabeel~Rehemtulla}
\Northwestern
\CIERA
\SkAI

\author[0000-0002-7015-3446]{Nicolas~Meza~Retamal}
\UCD

\author[0009-0006-3342-6181]{Olivia~M.~Robinson}
\UCSC

\author[0000-0002-7559-315X]{C\'{e}sar Rojas-Bravo}
\UCSC

\author[0000-0002-6688-0800]{Devendra~K.~Sahu}
\IIA

\author[0000-0003-4102-380X]{David~J.~Sand}
\UA

\author[0000-0002-8538-9195]{Brian~P.~Schmidt}
\ANU

\author[0000-0001-6797-1889]{Steve Schulze}
\CIERA

\author[0009-0002-5096-1689]{Michaela Schwab}
\Rutgers

\author[0000-0002-4022-1874]{Manisha Shrestha}
\UA

\author[0000-0003-2445-3891]{Matthew~R.~Siebert}
\STScI

\author[0000-0003-3801-1496]{Sunil Simha}
\CIERA
\UChicago

\author[0000-0001-5510-2424]{Nathan Smith}
\UA

\author[0000-0003-1546-6615]{Jesper Sollerman}
\OKCAstro

% \author[0000-0003-4524-6883]{Shubham~Srivastav}
% \Oxford

\author[0000-0001-8073-8731]{Bhagya~M.~Subrayan}
\UA

\author[0000-0003-4610-1117]{Tam\'as Szalai}
\USzeged
\MTAELTE

\author[0000-0002-5748-4558]{Kirsty Taggart}
\UCSC

\author[0000-0002-0525-0872]{Rishabh Singh Teja}
\IIA

\author[0000-0001-7380-3144]{Tea Temim}
\Princeton

\author[0000-0001-9834-3439]{Jacco~H.~Terwel}
\UDublin
\NOT

\author[0000-0002-1481-4676]{Samaporn Tinyanont}
\Thailand

\author[0000-0001-8818-0795]{Stefano Valenti}
\UCD

\author[0000-0001-9051-1338]{Jorge Anais Vilchez}
\CITEVA

\author[0000-0001-8764-7832]{J\'{o}zsef Vink\'{o}}
\Konkoly
\USzeged
\ELTE
\UT

\author[0009-0003-8229-0127]{Aya L. Westerling}
\UCSC

\author[0000-0002-6535-8500]{Yi Yang}
\Tsinghua
\UCB

\author[0000-0002-2636-6508]{WeiKang Zheng}
\UCB

\begin{abstract}
We present panchromatic optical $+$ near-infrared (NIR) $+$ mid-infrared (MIR) observations of the intermediate-luminosity Type Iax supernova (SN~Iax) 2024pxl and the extremely low-luminosity SN~Iax 2024vjm. \textit{JWST} observations provide unprecedented MIR spectroscopy of SN~Iax, spanning from $+$11 to $+$42~days past maximum light. We detect forbidden emission lines in the MIR at these early times while the optical and NIR are dominated by permitted lines with an absorption component. Panchromatic spectra at early times can thus simultaneously show nebular and photospheric lines, probing both inner and outer layers of the ejecta. We identify spectral lines not seen before in SN~Iax, including [\ion{Mg}{2}]~4.76\um, [\ion{Mg}{2}]~9.71\um, [\ion{Ne}{2}]~12.81\um, and isolated \ion{O}{1}~2.76\um\ that traces unburned material. Forbidden emission lines of all species are centrally peaked with similar kinematic distributions, indicating that the ejecta are well mixed in both SN~2024pxl and SN~2024vjm, a hallmark of pure deflagration explosion models. Radiative transfer modeling of SN~2024pxl shows good agreement with a weak deflagration of a near-Chandrasekhar-mass white dwarf, but additional IR flux is needed to match the observations, potentially attributable to a surviving remnant. Similarly, we find SN~2024vjm is also best explained by a weak deflagration model, despite the large difference in luminosity between the two supernovae. Future modeling should push to even weaker explosions and include the contribution of a bound remnant. Our observations demonstrate the diagnostic power of panchromatic spectroscopy for unveiling explosion physics in thermonuclear supernovae.
\end{abstract}


\section{Introduction \label{sec:intro}}
Type Iax supernovae \citep[SN Iax;][]{Foley2013}, also known as 02cx-like SN~Ia after the class prototype SN~2002cx \citep{Filippenko2003, Li2003}, are a peculiar subclass of thermonuclear SN that are characterized by lower kinetic energies, luminosities, and velocities than the ``normal" Type Ia supernovae (SN~Ia) used in cosmological analyses \citep[see][for a review]{Jha2017}. SN~Iax exhibit a greater diversity in observational properties than normal SN~Ia, with luminosities spanning over a factor 100 range in brightness, from $M_r=-12.7$~mag for SN~2021fcg \citep{Karambelkar2021} to $M_B=-18.3$~mag for SN~2012Z \citep{Stritzinger2015}, and line velocities at peak ranging from $\sim$2000 to $\sim$7000\kms\ \citep[e.g.,][]{Jha2017}.

% SWJ: do a search for some other recent Iax reference.. Barna et al., others? We want to encourage people (and especially non-collaborators) to keep working on Iax

% SWJ: there was an update paper to Fink et al. 2014 with some more models -- let's cite that below and maybe consider running RT through their weaker ones while paper is being reviewed
% SWJ: look also for anything from Amir Michaelis who presented on a ``failed detonation'' scenario for Iax at Padova and Garching

A leading explosion model for SN~Iax is the weak deflagration of a white dwarf (WD) with a near-Chandrasekhar mass ($M_\mathrm{Ch}$), where the explosion is not powerful enough to fully unbind the star and a remnant is left behind \citep{Branch2004, Jordan2012, Kromer2013, Fink2014}. These models have been shown to match the observed light curves and spectroscopic evolution of relatively high-luminosity SN~Iax quite well \cite[e.g.,][]{Magee2016,Camacho-Neves2023, Maguire2023}. SN~2012Z, a SN~Iax for which a He star companion was detected in pre-explosion imaging \citep{McCully2014}, remains visible with a flux excess above the pre-explosion flux over 2500~days later \citep{McCully2022, Schwab2025}, potentially signifying continuing emission from a bound remnant. Furthermore, \cite{Maeda2022} found evidence in the $\sim$500~day spectrum of the intermediate-luminosity SN Iax 2019muj for an Fe-rich innermost ejecta component that may have been ejected from the remnant via energy input from radioactive decay of iron-group elements (IGEs; similar to the remnant-driven wind theory from \citealt{Shen2017}).

However, it is unclear whether deflagration models and He-star companions can account for the heterogeneity of SN~Iax. \cite{Camacho-Neves2023} find that TARDIS radiative transfer models based on the lowest-energy deflagration model from \cite{Fink2014} match observations of SN~2014dt very well over the course of hundreds of days. However, SN~2014dt is a fairly luminous SN~Iax, making it problematic to explain extremely low-luminosity SN~Iax such as SN~2008ha \citep{Valenti2009,Foley2009}, SN~2010ae \citep{Stritzinger2014}, SN~2019gsc \citep{Tomasella2020}, SN~2020kyg \citep{Srivastav2022, Singh2023}, and SN~2021fcg \citep{Karambelkar2021}. \cite{Lach2022} explore a wide parameter space with their models, extending the brightness range somewhat lower, but not enough to explain the faintest objects. \cite{Kromer2015} proposed the deflagration of a hybrid C/O/Ne WD \citep{Meng2014} as method of producing extremely low-luminosity SN Iax. This model creates a transient with similar luminosity to SN~2008ha, but its light curve evolution is too rapid. \cite{Feldman2023b} discuss that simulating low-luminosity SN~Iax events is difficult due to their sparse, slow, and cool ejecta, which are highly sensitive to the simulation parameters used.

Recently, \cite{Callan2024} introduced the effects of a luminous central source (i.e., a remnant polluted by $^{56}$Ni deflagration ashes; \citealt{Shen2017}) into radiative transfer calculations of model light curves and spectroscopy. They found that the contribution of the remnant emission produces a slower post-maximum light curve decline and enhanced spectral continuum flux, improving the agreement between models and the data, especially for intermediate- and low-luminosity models.

SN~Iax exhibit the unique characteristic that their optical spectra remain dominated by permitted line transitions displaying a P-Cygni profile (for $>$500~days in SN~2014dt;  \citealt{Camacho-Neves2023}), indicating that densities stay high and an optically thick photosphere persists to very late times \citep{Jha2006, Jha2017}. \cite{Camacho-Neves2023} suggest that these permitted lines at late times may arise from a remnant-driven optically thick wind \citep{Shen2017}. Forbidden line emission, which signifies the onset of the nebular phase in normal SN~Ia at around 100~days post maximum in the optical, do emerge in SN~Iax over time; however, they coexist with permitted lines, whereas the permitted lines disappear in normal SN~Ia. In low-luminosity SN~Iax, where the ejecta masses are lower and the ejecta turn optically thin more quickly, forbidden lines emerge very early \citep[at $\sim+30$~days post maximum in SN~2020kyg;][]{Singh2023}.

Traditionally, SN are observed over time to probe the SN ejecta structure from outer to interior layers as the photosphere gradually recedes in the ejecta co-moving frame, eventually becoming optically thin in all layers in the nebular phase (although we note that some ultraviolet (UV) transitions remain optically thick even at late epochs). Panchromatic spectra over a large wavelength range (e.g., UV to IR), however, can reveal different layers of the ejecta at the same time because the optical depth is frequency dependent \citep{Pinto2000}. At MIR wavelengths, spectra are expected to become optically thin much earlier, allowing us to probe the interior layers in the MIR while the optical flux still emerges from higher-velocity outer layers.

Recent \textit{JWST} observations in the MIR of both normal and peculiar SN~Ia provide a new probe of progenitor and explosion models \citep{Kwok2023, DerKacy2023, Chen2023, Siebert2024, Kwok2024, DerKacy2024, Ashall2024}. Spectral lines in the near-infrared (NIR) and MIR are more isolated, making the contribution from individual lines easier to distinguish \citep[e.g.,][]{Pinto2000, Gerardy2007, Kwok2023}. Furthermore, while the optical and ground-based NIR wavelengths are dominated by emission lines from various ionization states of mainly Fe, longer wavelengths ($\lambda>2.5$\um) host strong lines of Ni and Co ions, as well as intermediate-mass elements (IMEs) such as Ne, S, Ar, and Ca.

Here, we present the first MIR spectra of SN~Iax, with \textit{JWST} observations of the intermediate-luminosity SN~Iax 2024pxl and the extremely low-luminosity SN~Iax 2024vjm. These spectra are also the earliest MIR spectra of any thermonuclear SN published to date, with phases ranging from $+11$ to $+42$~days post maximum light. SN~2024pxl and SN~2024vjm share significant NIR and MIR spectral similarities to each other, and distinct differences from the MIR spectra of other SN~Ia (SN~2021aefx, normal Ia, \citealt{Kwok2023, DerKacy2023, Ashall2024}; SN~2022xkq, 91bg-like, \citealt{DerKacy2024}; SN~2022pul, 03fg-like, \citealt{Siebert2023, Kwok2024}). These unique MIR spectral features may point to distinct differences in the way the WD explosion proceeds. For details on the discovery, photometric evolution, and ground-based optical$+$NIR spectral evolution of SN~2024pxl, see \cite{Singh2025} and \cite{Hoogendam2025}, and for SN~2024vjm, see \cite{Zimmerman2025}. In this work, we analyze the early panchromatic (optical$+$NIR$+$MIR) evolution of SN~2024pxl and SN~2024vjm with an emphasis on MIR wavelengths (hereafter $\lambda > 2.5$~\um).

\begin{figure*}
    \centering
    \includegraphics[width=\textwidth]{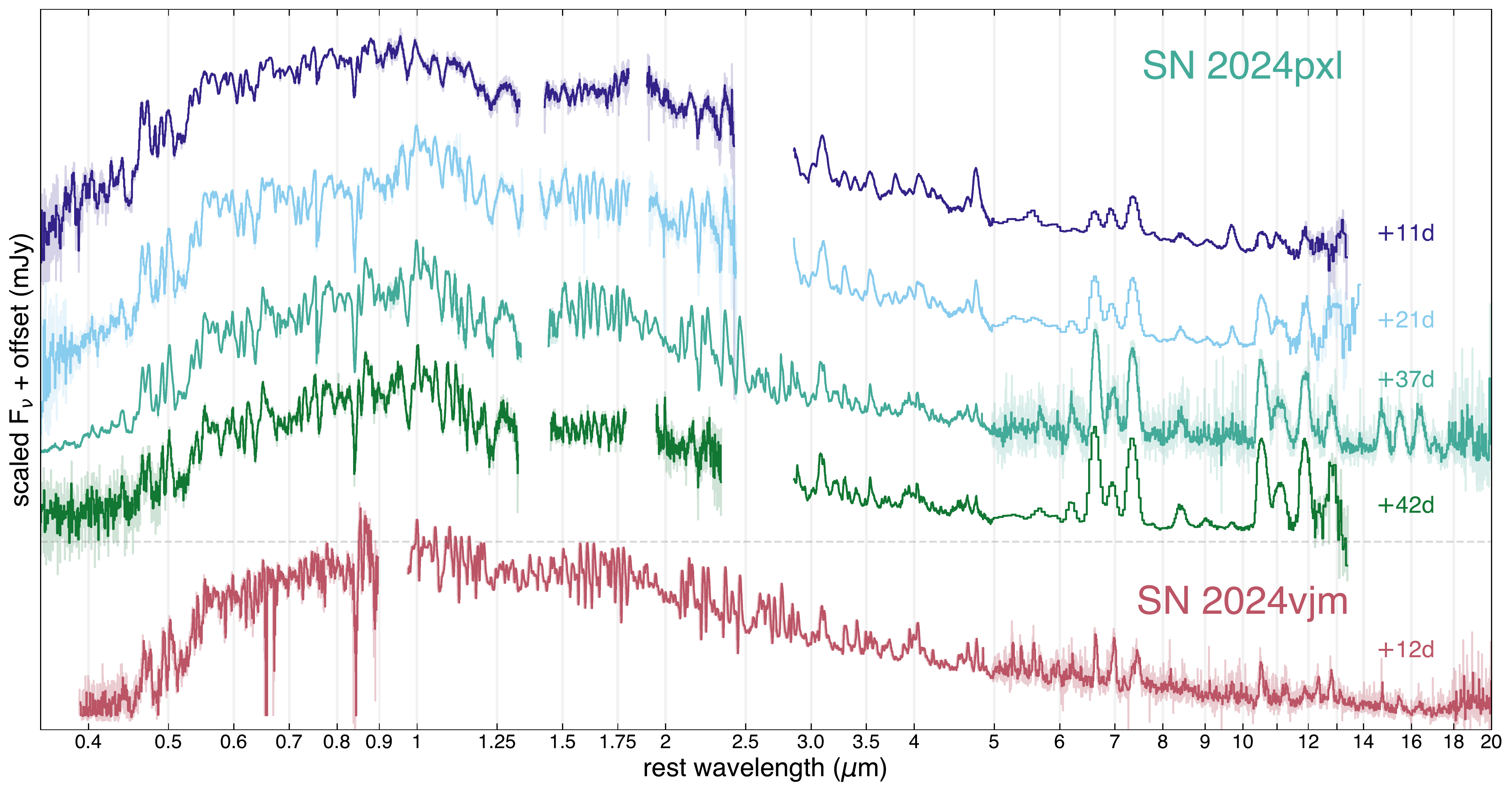}
    \caption{Panchromatic spectra of SN~2024pxl at $+$11, $+$21, $+$37, and $+$42~days post-\textit{B}$_\mathrm{max}$ compared to  SN~2024vjm at $+$12~days post-\textit{B}$_\mathrm{max}$. Flux density is shown in an arcsinh scaling for display purposes over a large range, and each spectrum is offset for visual clarity. Each  is a combination of ground-based optical and NIR spectra with NIR and MIR spectra from \textit{JWST} at similar phases. Details are given in \autoref{tab:obs}.}
    \label{fig:all_spec}
\end{figure*}

% \begin{figure*}
%     \centering
%     \includegraphics[width=\textwidth]{JWST_24pxl_24vjm.pdf}
%     \caption{Caption}
%     \label{fig:enter-label}
% \end{figure*}

\section{Observations \label{sec:obs}}

\subsection{JWST Data \label{sec:JWST_data}}
The \textit{JWST} data presented in this paper were obtained through three \textit{JWST} programs\footnote{All \textit{JWST} data is publicly available on MAST at DOI: \href{https://archive.stsci.edu/doi/resolve/resolve.html?doi=10.17909/vbq2-sg84}{10.17909/vbq2-sg84}}. Three epochs of NIR$+$MIR spectra of SN~2024pxl were collected as part of \textit{JWST}~GO~5232 (PI L.~A.~Kwok) using the medium resolution Near-Infrared Spectrograph (NIRSpec) with the Fixed Slit (FS) G395M grating and the Mid-Infrared Instrument (MIRI) with the Low Resolution Spectrograph (LRS) slit mode. These spectra were obtained on 13-Aug-2024, 23-Aug-2024, and 13-Sep-2024, corresponding to $+$11, $+$21, and $+$42~days post \textit{B}-band maximum-light, respectively. One NIR$+$MIR spectrum of SN~2024pxl was collected through \textit{JWST}~GO~6850 (PI S.~W.~Jha) using the NIRSpec FS G235M$+$G395M gratings and the MIRI Medium Resolution Spectrograph (MRS) on 08-Sep-2024 at $+$37~days post \textit{B}-band maximum-light. 

Our \textit{JWST} observation of SN~2024vjm was executed through \textit{JWST}~DD~6811 (PI L.~A.~Kwok) on 01-Oct-2024 at $+$12~days post \textit{B}-band maximum-light using the NIRSpec G140M$+$G235M$+$G395M gratings and MIRI MRS.

All NIRSpec grating data (G140M, G235M, and G395M) were reduced using the standard automatic \textit{JWST} pipeline as obtained on the Mikulski Archive for Space Telescopes (MAST). Similarly, all MIRI/LRS data were reduced through the automatic, standard \textit{JWST} pipeline (available on MAST). 

%We check that the LRS flux is well-calibrated by comparing F1000W synthetic photometry from the spectrum to measured photometry from the LRS verification image and find agreement within XX\%. 

However, in comparing our LRS data of SN~2024pxl (at $+11$, $+21$, and $+42$~days), to the MRS data of SN~2024pxl (at $+37$~days), we find discrepancies between the wavelengths of features $<7.5$\um. The LRS wavelength calibration has been known to be inaccurate at the shortest wavelengths and efforts have been made to address this issue; however, we find that the wavelength solution is still not precise and suggest future calibration observations of sources with clear spectral features at $<$7.5\um\ be taken contemporaneously with MRS and LRS in order to correct the LRS wavelength calibration to the MRS wavelength calibration. 

In this work, we calibrate our $+$42~day LRS spectrum to the $+$37~day MRS spectrum of SN~2024pxl, as there is minimal velocity evolution of the SN emission lines within these 5 days. We then apply this correction to the $+$11 and $+$21~day spectra of SN~2024pxl. Additional details of our LRS wavelength calibration correction are given in \autoref{sec:LRS_cal}.

The MIRI/MRS observations of SN~2024pxl and SN~2024vjm were reduced through the same procedure, as follows. First, we re-process the stage2 \texttt{*rate.fits} files through the \textit{JWST} pipeline to produce a single aligned data cube containing Channels 1$+$2$+$3$+$4, using Section 3 of a Github notebook developed by M. Shahbandeh\footnote{ \url{https://github.com/shahbandeh/MIRI_MRS/blob/main/MRS_reductions.ipynb}}. Next, instead of proceeding to the background subtraction section of this notebook, we use the Github AstroBkgInterp notebook developed by B. Nickson\footnote{ \url{https://github.com/brynickson/AstroBkgInterp}}. This background subtraction notebook is optimal for interpolating around the background of a point source and subtracting it off. Fortunately, SN~2024pxl and SN~2024vjm do not have complicated MRS backgrounds and this routine works very well to isolate the SN flux. We choose to truncate the MRS spectra of these faint objects at 20\um, as the noise in MRS/Channel~4 becomes too large for weak emission lines at these longest wavelengths to be detected.

\subsection{Ground-based Optical and NIR Data}
All five epochs of \textit{JWST} spectroscopy presented here are greatly enhanced by contemporaneous ground-based optical and NIR observations, allowing us to construct panchromatic optical $+$ NIR $+$ MIR spectra (see \autoref{tab:obs}). \autoref{fig:all_spec} shows these panchromatic spectra which span the wavelength range 0.35\um\--20\um. \cite{Singh2025} provide details on these optical and NIR spectra. Some of the data come from the Global Supernova Project (GSP).

For the \textit{JWST} epochs of SN~2024pxl taken with NIRSpec/G395M $+$ MIRI/LRS, there is a wavelength coverage gap between the ground-based NIR spectra and NIRSpec/G395M. We mangle the optical spectra to photometry, and scale the NIR spectra to the optical spectrum. Without NIR photometry, we are unable to mangle the NIR spectra so the absolute flux calibration of the ground-based NIR spectra is uncertain; none of the analysis presented here depends sensitively on this.

The NIRSpec/G235M$+$G395M $+$ MIRI/MRS observation of SN~2024pxl at $+$37~days overlaps with the optical$+$NIR VLT/XSHOOTER spectrum, so we scale the VLT data to match the flux level of the \textit{JWST} observations and splice the two spectra together at 1.7\um\ where the {\it JWST} data start.

For SN~2024vjm, we scale the optical spectra to photometry and combine and scale another optical spectrum from a few days later with better red coverage to help span the wavelength gap between where the NIRSpec/G140M$+$G235M$+$G395M spectrum starts. Visually, these spectra appear in close agreement, so we do not apply any further flux scaling.

\begin{table*}
    \centering
    \begin{tabular}{llllll}
        \hline
        \textbf{Date} & \textbf{MJD} & \textbf{Phase} (d) & \textbf{Telescope / Instrument} & \textbf{Wavelength range}  & \textbf{Program / PI}\\
        \hline
        \hline
        \multicolumn{6}{l}{\textbf{SN~2024pxl Epoch 1}} \\
        % \hline
        2024-08-13 & 60535.4 & $+$11.3 & JWST / NIRSpec / G395M & 2.9 -- 5\um & JWST-GO-5232 / L.~A.~Kwok \\
        2024-08-13 & 60535.4 & $+$11.3 & JWST / MIRI / LRS & 5 -- 14\um & JWST-GO-5232 / L.~A.~Kwok \\
        2024-08-13 & 60535.6 & $+$11.5 & HCT / HFOSC & 3900 -- 9000~\AA\ & HCT-2024-C3-P22 / D. K. Sahu \\
        2024-08-14 & 60537.3 & $+$12.7 & SALT / RSS & 3500 -- 9000~\AA\ & 2024-1-MLT-004 / L.~A.~Kwok \\
        2024-08-14 & 60537.4 & $+$12.8 & NOT / ALFOSC & 3500 -- 9000~\AA\ & 68-501 / J.~Sollerman \\
        2024-08-14 & 60537.3 & $+$12.2 & FTN / FLOYDS & 3500 -- 9300~\AA\ & Global Supernova Project / D.~A.~Howell\\
        2024-08-14 & 60537.4 & $+$12.8 & Lick / Kast & 3500 -- 9300~\AA\ & 5474 / A.~V.~Filippenko \\
        2024-08-12 & 60534.9 & $+$10.3 & IRTF / SpeX & 0.92 -- 2.4\um\ & 2024B078 / A.~P.~Ravi \\
        % \hline
        \hline
        \multicolumn{6}{l}{\textbf{SN~2024pxl Epoch 2}} \\
        % \hline
        2024-08-23 & 60545.0 & $+$20.9 & JWST / NIRSpec / G395M & 2.9 -- 5\um & JWST-GO-5232 / L.~A.~Kwok \\
        2024-08-23 & 60544.9 & $+$20.8 & JWST / MIRI / LRS & 5 -- 14\um & JWST-GO-5232 / L.~A.~Kwok \\
        2024-08-26 & 59548.0 & $+$23.4 & SOAR / TripleSpec & 0.9 -- 2.5\um\ & SOAR2024B-006 / A.~P.~Ravi \\
        2024-08-23 & 60545.8 & $+$21.2 & FTN / FLOYDS & 3500 -- 9300~\AA\ & Global Supernova Project / D.~A.~Howell \\
        2023-08-23 & 60546.4 & $+$21.8 & NOT / ALFOSC & 3500 -- 9000~\AA\ & 68-501 / J.~Sollerman \\
        2024-08-23 & 60545.9 & $+$21.8 & SALT / RSS & 3500 -- 9000~\AA\ & 2024-1-MLT-004 / L.~A.~Kwok \\
        2024-08-23 & 60546.1 & $+$21.49 & HCT / HFOSC & 3900 -- 9000~\AA\ & HCT-2024-C3-P22 / D. K. Sahu \\
        % \hline
        \hline
        \multicolumn{6}{l}{\textbf{SN~2024pxl Epoch 3}} \\
        % \hline
        2024-09-08 & 60561.4 & $+$37.3 & JWST / NIRSpec / G235M & 1.7 -- 2.9\um & JWST-GO-6580 / S.~W.~Jha \\
        2024-09-08 & 60561.4 & $+$37.3 & JWST / NIRSpec / G395M & 2.9 -- 5\um & JWST-GO-6580 / S.~W.~Jha \\
        2024-09-08 & 60561.2 & $+$37.1 & JWST / MIRI / MRS & 5 -- 28\um & JWST-GO-6580 / S.~W.~Jha \\
        2024-09-05 & 60558.0 & $+$33.9 & VLT / XSHOOTER & 0.3 --2.5\um\ & 114.27JL.001 / S.~Blondin \\
        2024-09-08 & 60562.5 & $+$37.0 & MMT / Binospec & 3500 -- 9000~\AA\ & UAO-G200-24B / A.~A.~Miller \\
        2024-09-08 & 60561.7 & $+$37.1  & Lick / Kast & 3500 -- 9300~\AA\ & 2024B-S022 / R.~J.~Foley \\
        \hline
        % \hline
        \multicolumn{6}{l}{\textbf{SN~2024pxl Epoch 4}} \\
        % \hline
        2024-09-13 & 60566.3 & $+$42.2 & JWST / NIRSpec / G395M & 2.9 -- 5\um & JWST-GO-5232/L.~A.~Kwok \\
        2024-09-13 & 60566.3 & $+$42.2 & JWST / MIRI / LRS & 5 -- 14\um & JWST-GO-5232 / L.~A.~Kwok \\
        2024-09-13 & 60566.6 & $+$42.5 & GTC / EMIR YJ$+$HK & 0.93 -- 2.4\um\ & GTCMULTIPLE2A-24B / L.~Galbany\\
        2024-09-13 & 60566.7 & $+$42.6 & Lick / Kast & 3500--9300~\AA\ & 2024B-S022 / R.~J.~Foley \\
        2024-09-14 & 60567.3 & $+$43.2 & FTN / FLOYDS & 3500 -- 9300~\AA\ & Global Supernova Project / D.~A.~Howell \\
        2024-09-14 & 60567.4 & $+$43.3 & ANU / WiFeS & 3500 -- 9300~\AA\ & Australian Supernova Alliance \\
        \hline
        % \hline
        \multicolumn{6}{l}{\textbf{SN~2024vjm}} \\
        % \hline
        2024-10-01 & 60584.6 & $+$12 & JWST / NIRSpec / G140M & 0.97 -- 1.7\um & DD 6811 | L.~A.~Kwok \\
        2024-10-01 & 60584.6 & $+$12 & JWST / NIRSpec / G235M & 1.7 -- 2.9\um & DD 6811 / L.~A.~Kwok \\
        2024-10-01 & 60584.6 & $+$12 & JWST / NIRSpec / G395M & 2.9 -- 5\um & DD 6811 / L.~A.~Kwok \\
        2024-10-02 & 60585.0 & $+$12 & JWST / MIRI / MRS & 5 -- 28\um & DD 6811 / L.~A.~Kwok \\
        2024-10-01 & 60584.8 & $+$12 & SALT / RSS  & 3500--8000~\AA\ & 2024-1-MLT-004 / L.~A.~Kwok \\
        2024-10-05 & 60588.8 & $+$16 & SALT / RSS  & 8000--9300~\AA\ & 2024-1-MLT-004 / L.~A.~Kwok \\
        \hline
    \end{tabular}
    \caption{Data used to create the panchromatic spectra in \autoref{fig:all_spec}.}
    \label{tab:obs}
\end{table*}

% \begin{figure*}
%     \centering
%     \includegraphics[width=\textwidth]{SN204pxl_JWST_evolution_small.pdf}
%     \caption{Caption}
%     \label{fig:enter-label}
% \end{figure*}

\section{Spectral Analysis \label{sec:spec_analysis}}
A distinctive characteristic of SN~Iax spectra are their narrow line widths ($\sim$2000--7000\kms\ near peak), resulting from low ejecta velocities. These narrow lines allow individual lines to be isolated more easily than in normal SN~Ia where, especially in the optical, the lines are blended. Our \textit{JWST} spectra of SN~2024pxl and SN~2024vjm reveal a plethora of spectral lines, arising from optically thick permitted and optically thin forbidden transitions, which have never been observed before. In this section, we present NIR and MIR line identifications for SN~2024pxl and SN~2024vjm, and qualitatively compare these SN to each other and \textit{JWST} observations of other SN~Ia.

\begin{figure*}
    \centering
    \includegraphics[width=\linewidth]{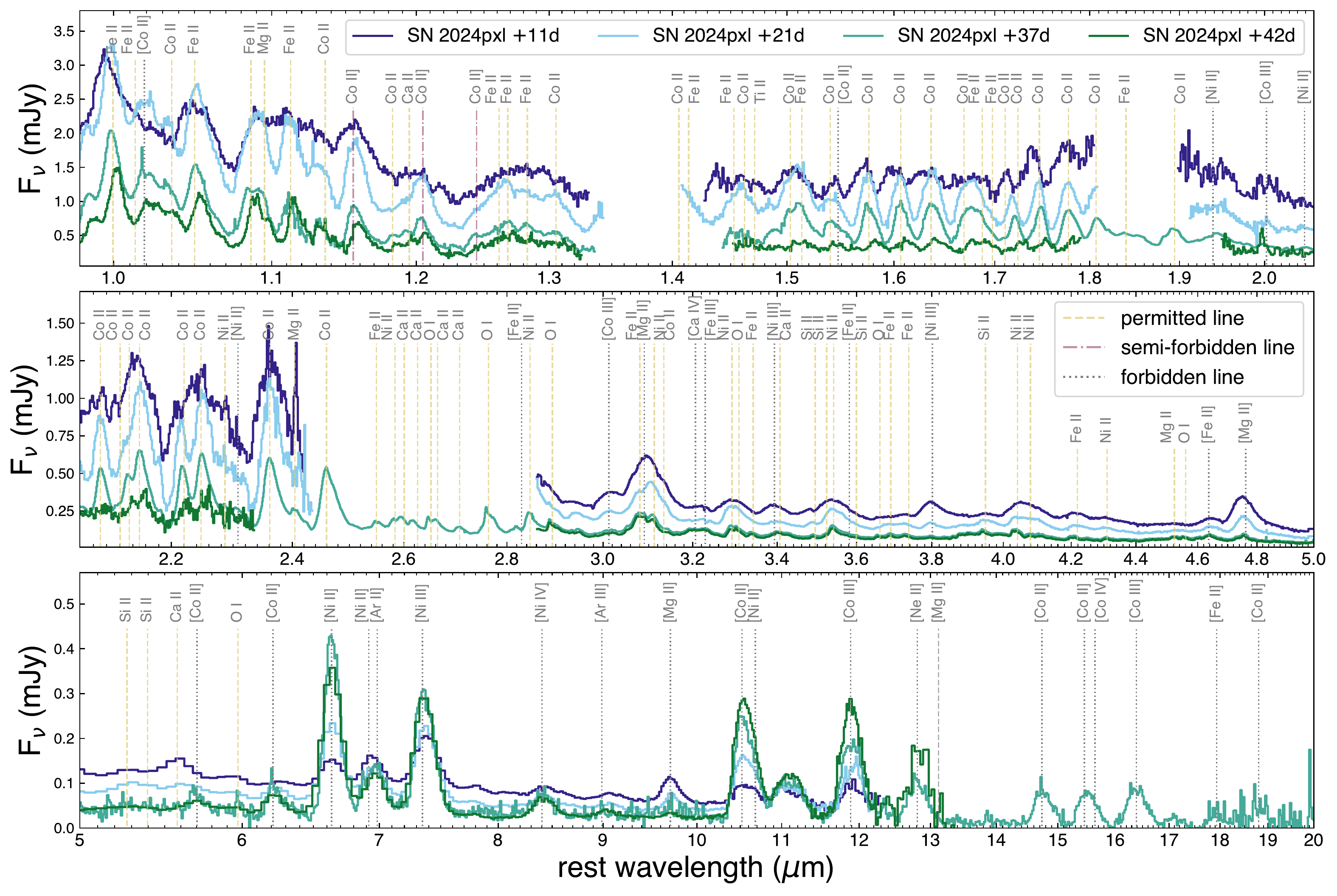}
    \caption{Line identifications for SN~2024pxl from 1.0 to 20\um. Permitted transitions are marked by yellow dashed lines, semiforbidden transitions are marked by purple dashed-dotted lines, and forbidden transitions are marked by gray dotted lines. Only the most dominant lines contributing to each feature are labeled.}
    \label{fig:24pxl_lineID}
\end{figure*}

\subsection{Line Identifications \label{sec:lineIDs}}
Previous observations of thermonuclear SN at wavelengths $\lambda > 2.5$\um\ at late times \citep{Gerardy2007, Kwok2023, DerKacy2023, Kwok2024, DerKacy2024} show strong nebular emission lines in the MIR corresponding to transitions to the ground-state or a low energy level. Line identifications from MIR models are also given by \cite{Blondin2023, DerKacy2024, Ashall2024}. We detect most of these MIR emission lines in our \textit{JWST} observations of SN~2024pxl and SN~2024vjm, as well as additional lines which may be unique to SN~Iax, and/or are due to the earlier phases of the observations. 

To aid in line identification of these complicated spectra, we use an atomic line list generated from a CMFGEN \citep{Hillier2012} radiative transfer simulation based on the lowest-energy pure deflagration model of \citet{Fink2014}. Information about this model is presented in \autoref{sec:models}, and details of our line identification procedure can be found in \autoref{sec:line_id_appendix}. These line identifications may also be helpful in securing uncertain optical line identifications.

\begin{figure*}
    \centering
    \includegraphics[width=\linewidth]{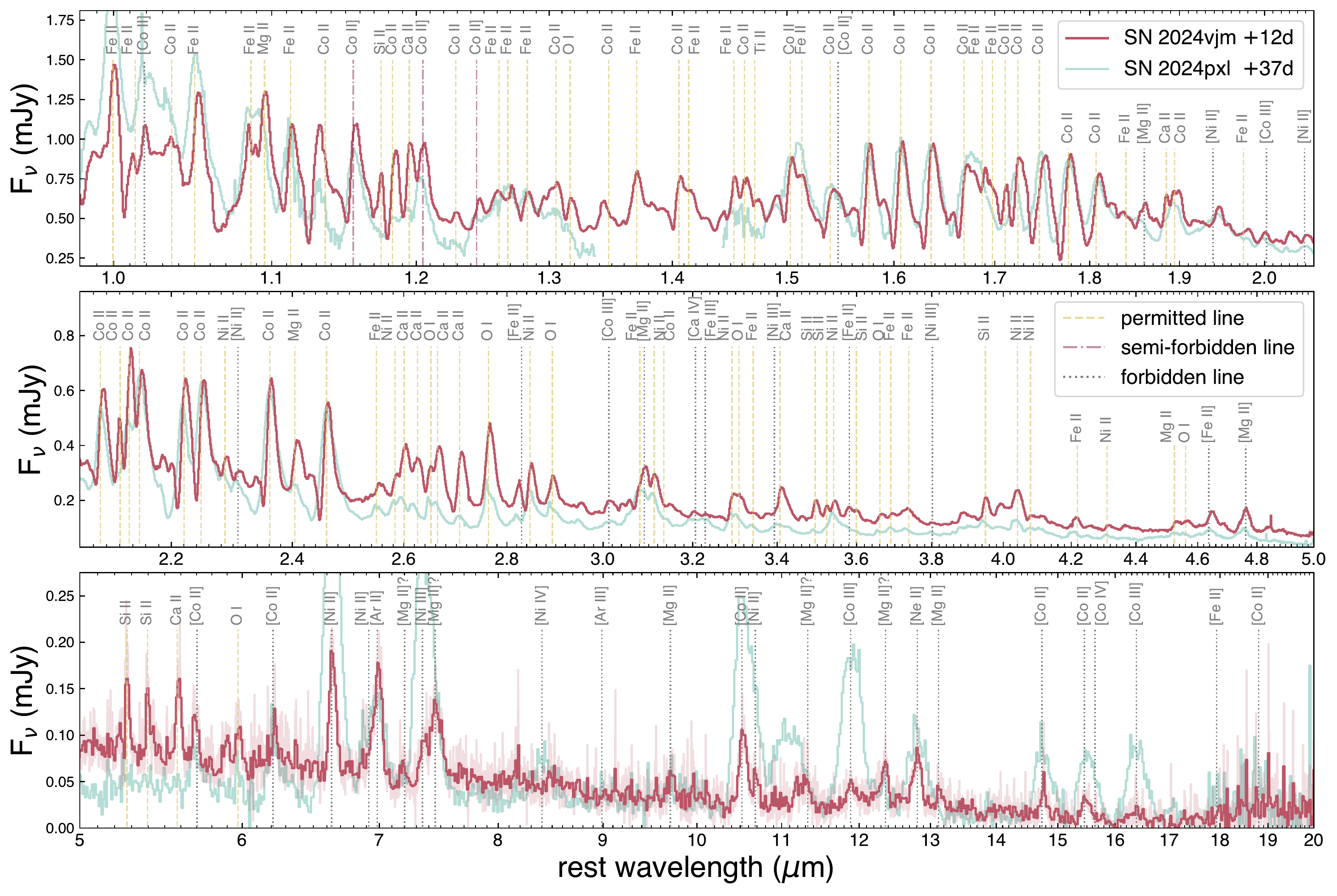}
    \caption{Line identifications for SN~2024vjm from 1.0 to 20\um. Permitted transitions are marked by yellow dashed lines, semiforbidden transitions are marked by purple dashed-dotted lines, and forbidden transitions are marked by gray dotted lines. Only the most dominant lines contributing to each feature are labeled. For comparison, SN~2024pxl at $+$37~days is plotted in light teal.}
    \label{fig:24vjm_lineID}
\end{figure*}

\autoref{fig:24pxl_lineID} and \autoref{fig:24vjm_lineID} show our line identifications for SN~2024pxl and SN~2024vjm, respectively. While some line identifications (especially weaker features or those at earlier epochs where velocities are higher and there is more line overlap) should be regarded as tentative, we find that most observed lines match well with our model line list. For clarity, in the case where many lines contribute to a given feature, we label only the dominant lines, determined by their relative Sobolev equivalent width (EW).

\subsubsection{NIR Co II lines}
Published NIR spectra are available for only a handful of SN~Iax \citep[SN~2005hk, SN~2008ge, SN~2010ae, SN~2012Z, SN~2014ck, SN~2015H, SN~2019muj, SN~2020udy][]{Kromer2013, Stritzinger2014, Stritzinger2015, Tomasella2016, Magee2016, Barna2021b,Maguire2023}, but they all display prominent permitted \ion{Co}{2} lines from 1.5--2.5\um\ when observed more than a week post peak. \cite{Stritzinger2014} find that these \ion{Co}{2} lines increase in prominence in low-luminosity objects with the lowest velocities. SN~2024pxl and SN~2024vjm also display very clear \ion{Co}{2} lines. The \textit{JWST} data of both SN at these epochs are high S/N, and distinct P-Cygni profiles are detected. Normal SN~Ia also display these Co lines, however, no clear P-Cygni profiles are visible because the higher velocities cause them to overlap and blend together.

\subsubsection{O I, Mg II, Si II, and Ca II lines}
Our \textit{JWST} spectra of both SN~2024pxl and SN~2024vjm reveal and resolve several lines of permitted \ion{O}{1}, \ion{Ca}{2}, and \ion{Si}{2} in the 1.0--6.0\um\ range. In particular, the wavelength region from 2.5--3.0\um, accessible by \textit{JWST}, hosts a cluster of \ion{Ca}{2} and \ion{O}{1} lines. \ion{O}{1}~2.76\um\ is strong and isolated in both SN~2024pxl and SN~2024vjm; while some oxygen may be the result of C burning, much of the oxygen in SN~Iax is thought to be pre-existing, so this line may be an excellent tracer for unburned material in the ejecta. 

Permitted O, Mg, Si, and Ca lines are stronger (relatively) in SN~2024vjm than in SN~2024pxl, and several additional lines appear between 5--6\um\ in SN~2024vjm. The increased prominence of these lines may suggest that SN~2024vjm has a higher mass fraction of IMEs and/or be an effect of the overall ejecta ionization state. We searched for signatures of carbon in the NIR and MIR, but we do not find anything convincing.

\subsubsection{[Mg II] lines}
We discover prominent lines at 4.76 and 9.71\um\ in SN~2024pxl that fade rapidly over time and identify them as forbidden [\ion{Mg}{2}]. These [\ion{Mg}{2}] lines arise from transitions between highly-excited energy levels forming a recombination sequence, with predicted wavelengths at 9.71, 6.94, 4.76, 3.09, and 1.86\um. These lines fade with time due to the decreasing {\sc iii}$\rightarrow${\sc ii} recombination rate resulting from the dropping densities and temperatures in the line-forming region. We detect these expected lines at 9.71, 4.76, 3.09, and 1.86\um, although the 3.09\um\ line is affected by line overlap, and the 1.86\um\ line is too weak to detect in SN~2024pxl. The line at 6.94\um\ is dominated by, but may contaminate, the [\ion{Ar}{2}]~6.98\um\ line.

Several additional weak emission lines in the MIR spectrum of SN~2024vjm also match with other [\ion{Mg}{2}] lines from our model line-list that are expected to be extremely weak, and we do not find compelling alternatives. We mark these identifications as tentative in \autoref{fig:24vjm_lineID} with question marks. Improvements in explosion models, radiative transfer post-processing, and potentially MIR atomic data are needed.

\subsubsection{Fading Ni II and [Ni III] lines \label{sec:fading_neb_lines}}

We find several lines between 3--5\um\ that also drop in strength over time in SN~2024pxl. The strongest lines in our model list at these wavelengths correspond to permitted \ion{Ni}{2} and \ion{Fe}{2}, and forbidden [\ion{Ni}{3}] and [\ion{Co}{3}] lines. Permitted lines are expected to fade over time as the ejecta expand, but fading IGE forbidden lines are more surprising. Radioactive decay may partially contribute to the fading Ni lines; however, the forbidden emission from Ni lines at $>$5\um\ instead grows. These fading lines arise from transitions between excited states higher than the MIR ground-state transition forbidden lines. 
%For example, $n_u=5$ and $n_l=4$ (1.74--2.07~eV, corresponding to $\sim$\num{2.3e4}~K) for [\ion{Ni}{3}]~3.80\um, $n_u=6$ and $n_l=4$ (1.74--2.10~eV) for [\ion{Ni}{3}]~3.38\um, $n_u=15$ and $n_l=11$ (2.59--3.00~eV) for [\ion{Co}{3}]~3.01\um. 
These energies are lower than those of the [\ion{Mg}{2}] transitions above. These lines may fade over time due to their somewhat higher energies (assuming these lines are thermally excited), or it may be attributed to a shift in ionization state seen in the MIR where the ratio of [\ion{Ni}{2}]/[\ion{Ni}{3}] increases early on (see \autoref{sec:ion_ratios} for additional discussion).

We note that in this 3-5\um\ region, permitted and forbidden lines of \ion{Fe}{2} appear to coexist. The \ion{Ni}{2} and \ion{Co}{2} lines in this region are predominantly permitted, while the \ion{Ni}{3} and \ion{Co}{3} lines are predominantly forbidden. Comparison to detailed modeling may be able to place constraints on the densities present in the ejecta, but is beyond the scope of the current work.

\subsection{Spectral Comparisons}

\subsubsection{SN~2024pxl vs. SN~2024vjm}
SN~2024pxl is an intermediate-luminosity SN~Iax with a peak absolute magnitude $M_B=-$16.10~mag \citep{Singh2025}, while SN~2024vjm is $\sim$3~mag fainter at $M_B\sim-$13~mag. Despite their large difference in luminosity, their early panchromatic spectra display many similarities. As seen in \autoref{fig:all_spec}, the $+$12~day panchromatic spectrum of SN~2024vjm closely resembles the $+$37~day panchromatic spectrum of SN~2024pxl, particularly in the NIR region from 1--5\um. In fact, SN~2024vjm at $+$12~days is spectroscopically more similar to the $+$37~day spectrum of SN~2024pxl than it is to the $+$11~day spectrum of SN~2024pxl, indicating that the extremely low-luminosity SN~2024vjm transitions to the nebular phase more quickly. 

\begin{figure*}
    \centering
    \includegraphics[width=\textwidth]{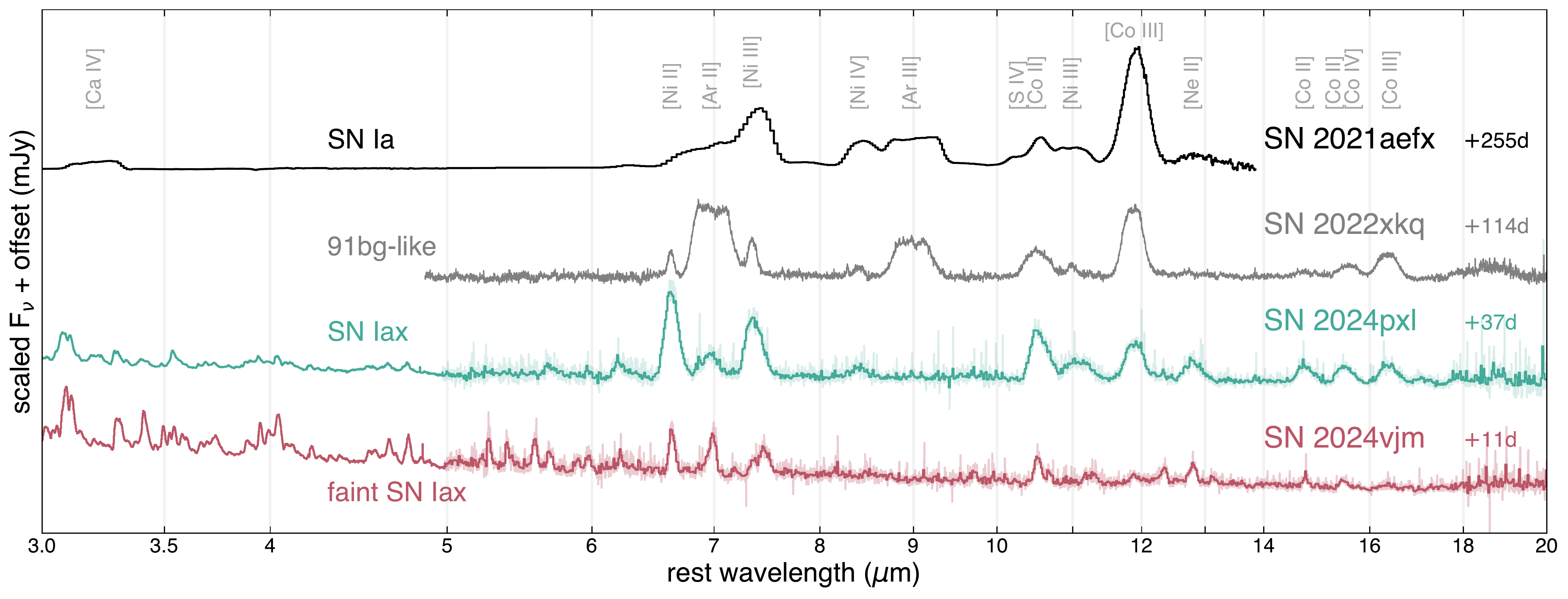}
    \caption{Comparison of the MIR spectra of normal SN~Ia~2021aefx \citep{Kwok2023}, 91bg-like SN~Ia~2022xkq \citep{DerKacy2024}, and the SN~Iax 2024pxl and 2024vjm. Notably, the SN~Iax IME profiles are centrally peaked indicating well-mixed ejecta, whereas SN~2021aefx and SN~2022xkq IME profiles (particularly [\ion{Ar}{2}]~6.89\um, and [\ion{Ar}{3}]~8.99\um) are flat-topped indicating stratified ejecta.}
    \label{fig:21aefx_22xkq_MIR_comp}
\end{figure*}

This is in line with observations of other low-luminosity SN~Iax which display emerging forbidden lines in their optical spectra as early as $\sim+30$~days \citep{Singh2023}. These objects likely transition to the nebular phase earlier due to lower specific kinetic energy ($KE/M_\mathrm{ej}$), with derived ejecta masses of $\sim$0.1--0.5\msun \citep[e.g.,][]{Foley2009, Valenti2009, Stritzinger2014, Tomasella2020, Srivastav2022, Singh2023, Karambelkar2021}.

Several differences between the spectra of SN~2024pxl and SN~2024vjm are apparent as well, the most obvious being that the lines are narrower in SN~2024vjm. This corresponds to lower velocities ($\sim$8000 vs. $\sim$3000\kms, respectively, for forbidden lines) and a narrower line-forming region, indicative of a steeper density profile or a density drop-off. Line velocity measurements are presented and discussed in \autoref{sec:phot_vels} and \autoref{sec:neb_vels}. The overall ionization state of SN~2024vjm is also lower than in SN~2024pxl. This is seen most clearly in the MIR wavelength region, where doubly and triply ionized species such as [\ion{Co}{3}], [\ion{Ar}{3}], [\ion{Ni}{4}], and [\ion{Co}{4}] are extremely weak or not detected. Finally, of particular interest, is the increased prominence in SN~2024vjm of low- and intermediate- mass elements (LMEs, IMEs) such as \ion{O}{1}, \ion{Mg}{2}, [\ion{Mg}{2}], \ion{Si}{2}, and \ion{Ca}{2}. This may point to an enhanced ejecta mass fraction of unburned material, LMEs, and IMEs in lower-luminosity objects.

\subsubsection{Comparison to SN~Ia}
\autoref{fig:21aefx_22xkq_MIR_comp} compares the MIR spectra of the normal SN~Ia 2021aefx \citep{Kwok2023} and the 91bg-like SN~Ia 2022xkq \citep{DerKacy2024} to the SN Iax MIR spectra. Note that in this work we re-reduce the MIRI/MRS spectrum of SN~2022xkq presented by \cite{DerKacy2024} (obtained through program JWST-GO-2114, PI~C.~Ashall, and publicly available on MAST) through the same process as the MRS spectra of SN~2024pxl and SN~2024vjm described in \autoref{sec:JWST_data}, which improves background subtraction. We caution that the phases of these observations of SN~2021aefx and SN~2022xkq ($+$255 and $+$114~days, respectively) are much later than the early phases of our SN~Iax observations. However, they are the closest comparisons currently available. Planned future late-time observations of SN~2024pxl through \textit{JWST}-GO-6580 will be a better comparison to the nebular MIR spectra of SN~2021aefx and SN~2022xkq.

Like all MIR spectroscopy of thermonuclear SN to date, SN~2024pxl and SN~2024vjm exhibit forbidden-line emission from ground-state transitions such as [\ion{Ni}{2}]~6.64\um, [\ion{Ar}{2}]~6.98\um, [\ion{Ni}{3}]~7.35\um, [\ion{Co}{2}]~10.52\um, and [\ion{Co}{3}]~11.89\um. In contrast to SN~2021aefx and SN~2022xkq, SN~2024pxl and SN~2024vjm lack strong emission from ions with high ionization energy such as [\ion{Ca}{4}], [\ion{Ar}{3}], and [\ion{S}{4}]. This is even more pronounced in SN~2024vjm where [\ion{Ar}{3}]~8.98\um, [\ion{Ni}{4}]~8.41\um, [\ion{Ni}{3}]~11.0\um, and [\ion{Co}{3}]~11.89\um\ are extremely weak, or not detected. 

The SN~Iax spectra also differ from SN~2021aefx and SN~2022xkq in that they exhibit [\ion{Ne}{2}]~12.81\um\ emission. \cite{Kwok2024} detect strong, broad, centrally-peaked [\ion{Ne}{2}]~12.81\um\ in SN~2022pul and interpret it as a signature of a violent merger, with the caveat noted by \citet{Blondin2023} that pure deflagrations can also mix Ne into the central ejecta. In the context of other observables for SN~2022pul, a pure deflagration was disfavored. Here, however, the centrally-peaked [\ion{Ne}{2}]~12.81\um\ line is much narrower, and given other contextual SN properties, we find that its presence points toward a pure deflagration model, further discussed in \autoref{sec:models}.

\begin{figure*}
    \centering
    \includegraphics[width=\linewidth]{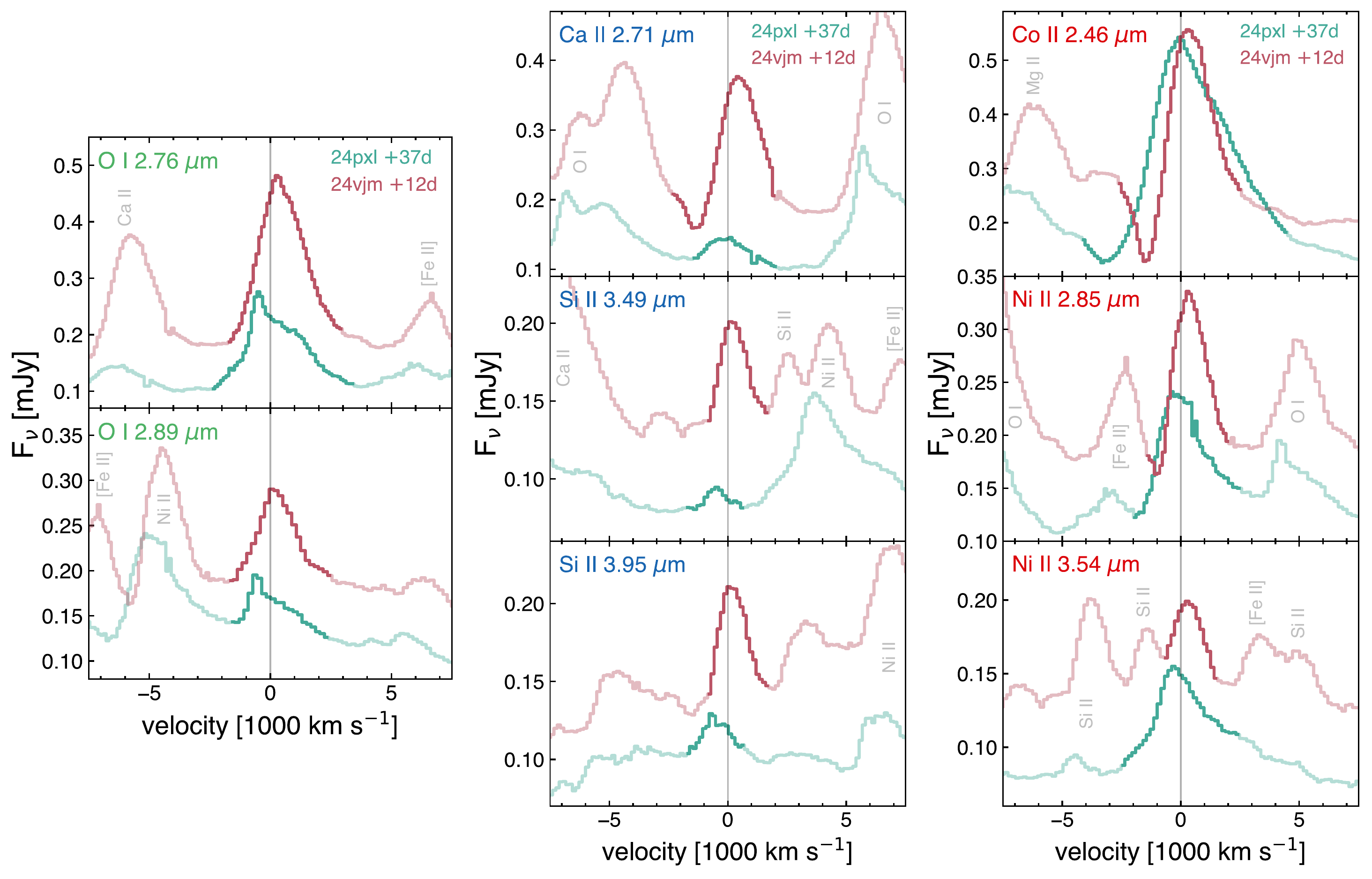}
    \caption{Comparison of selected relatively isolated permitted lines, some of which display a clear P-Cygni profile, in SN~2024pxl at $+$37~days (teal) and SN~2024vjm (pink) at $+$12~days. The features that are confidently associated with the labeled are given in full opacity. Regions that may be contaminated by other lines are shown in low opacity. The spectra have not been scaled or offset.}
    \label{fig:phot_profiles}
\end{figure*}

The flat-topped line profiles (originating from a thick shell emitting geometry) of the IMEs in SN~2021aefx and SN~2022xkq, specifically the [\ion{Ar}{2}]~6.98\um, [\ion{Ar}{3}]~8.98\um, and [\ion{Ca}{4}]~3.26\um\ lines, are hallmarks of explosions involving detonations, such as delayed- or double-detonations \citep{Kwok2023, DerKacy2023, Blondin2023, Kwok2024, DerKacy2024, Ashall2024}. In SN~2024pxl and SN~2024vjm, we observe that forbidden emission lines of both IGEs (e.g., [\ion{Ni}{2}], [\ion{Co}{3}], etc.) and IMEs (e.g., [\ion{Ar}{2}], [\ion{Ne}{2}], [\ion{Mg}{2}]) are centrally peaked. This is not just an effect of the differing phase -- as more interior layers are revealed over time, we anticipate that the line profiles will only become more peaked (assuming the ionization state does not change significantly). This indicates that the ejecta are mixed, rather than stratified, consistent with pure deflagration or violent merger models (see \autoref{sec:models}).

SN~2021aefx displays the broadest features while SN~2024vjm has the narrowest. At peak absolute magnitude $M_B=-18.01$~mag, SN~2022xkq is a member of the subluminous 91bg-like SN~Ia subclass \citep{Pearson2024}, making it comparable in luminosity to high-luminosity SN~Iax. SN~2022xkq and SN~2024pxl have roughly similar Co line widths, but SN~2022xkq has broader Ar emission, and narrower Ni emission. A fairer comparison of line widths between objects will be possible with later time \textit{JWST} observations of SN~2024pxl that will be more similar in phase.

\section{Permitted Lines \label{sec:phot}}

At the early phases of our \textit{JWST} observations of SN~2024pxl and SN~2024vjm, the optical spectra are still dominated by optically thick permitted lines displaying a clear absorption component (P-Cygni profile) characteristic of the photospheric phase. From around 1--2.5\um, the NIR is also dominated by optically thick permitted lines, mainly \ion{Fe}{2} and \ion{Co}{2}. Between 2.5--6\um, a transition occurs where forbidden lines begin to emerge, coexisting with permitted lines. In this wavelength range, some permitted lines, such as \ion{O}{1}~2.76\um, do not appear to display an absorption component, indicating they may be optically thin. At wavelengths longer than $\sim$6\um, the spectrum instead becomes dominated by forbidden emission. \autoref{fig:phot_profiles} compares the line profiles of SN~2024pxl at $+$37~days and SN~2024vjm at $+$12~days, in velocity space, of some selected, relatively isolated permitted lines.

\begin{figure}
    \centering
    \includegraphics[width=\linewidth]{unburned_OI_revised.pdf}
    \caption{Qualitative schematic showing that the \ion{O}{1} line morphology in SN~2024pxl may reflect predictions from \cite{Fink2014} of weak one-sided deflagrations. The base \ion{O}{1} emission (yellow, dotted lines) likely arises from the inner O-rich region ($<$5000\kms). The unburned channel (magenta, dashed lines) results from incomplete burning, and may correspond to the off-center narrow peak of the \ion{O}{1} emission. Right panel adapted from the N20def $^{16}$O of Figure 9 from \cite{Fink2014}.}
    \label{fig:unburned_OI}
\end{figure}

\subsection{O I Line Profiles}
Zooming in on the line profiles of \ion{O}{1}~2.76\um\ and \ion{O}{1}~2.89\um\ (\autoref{fig:phot_profiles}, \textit{left}), we find that these lines have similar widths between SN~2024pxl and SN~2024vjm (albeit at different phases), but the peaks are blueshifted and redshifted, respectively, by $\sim$500\kms. Within each SN, the \ion{O}{1} lines are consistent in shape, width, and kinematic offset. These permitted \ion{O}{1} lines do not display a P-Cygni profile, and do not appear to be significantly contaminated by lines where a blueshifted absorption component might be expected, indicating that these lines are likely optically thin. In SN~2024vjm, the \ion{O}{1} line morphology is centrally peaked and relatively symmetric, similar to the morphology of lines from other ions. The \ion{O}{1} lines in SN~2024pxl, however, have an asymmetric morphology with a narrow blueshifted peak and a broad red shoulder. 

This type of morphology can arise from an ejecta geometry with large-scale clumping, a single massive blob, or a unipolar jet \citep{Taubenberger2009}. The distribution of oxygen in thermonuclear SN is particularly important because it traces unburned material, although simulations indicate that a non-dominant fraction of it can also be a product of C-burning. \cite{Fink2014} find that their 3D hydrodynamical deflagration models with low numbers of ignition points can cause a one-sided deflagration that extinguishes before the burning wraps around the whole star, leaving behind a channel of unburned material. Furthermore, in their models with fewer than 10 ignition points, an asymmetric shell-like structure is imprinted onto the ejecta structure from the shock generated when the rebounding core hits outer layers that are still falling inward after the deflagration ceases and the bound parts contract. This unburned channel and shell can be best seen in the N20def model in their Figure 9 \citep{Fink2014}, but they note this channel is also present in several other models with $\lesssim$40 ignition points.

We suggest that the \ion{O}{1} line morphology in SN~2024pxl could reflect this predicted unburned structure, with a broader base emission component representing the inner O-rich region ($<$5000\kms), and the narrow component representing the channel. The shell structure is likely at too high velocity ($\sim$5000 to 10,000\kms) and low density to be detected in the emission profile. A qualitative schematic of this is shown in \autoref{fig:unburned_OI}. We note that the single ignition point N1def model compared to later in this work (see \autoref{sec:models}) does not show this unburned channel structure \citep{Fink2014}. However, the bright, moderately bright, and faint models of \cite{Lach2022} shown in their Figures 6 and 7 each exhibit a channel structure of unburned material. If the narrow blue peak seen in the \ion{O}{1} line emission in SN~2024pxl is indeed indicative of an unburned channel of material, then future observations of this line in SN~Iax may show narrow components at various kinematic offset, depending on the location of the unburned channel along the line-of-sight. The line profile of SN~2024vjm suggests that its explosion may have been more symmetric, or could be an effect of viewing angle.

\begin{figure*}
    \centering
    \includegraphics[width=\linewidth]{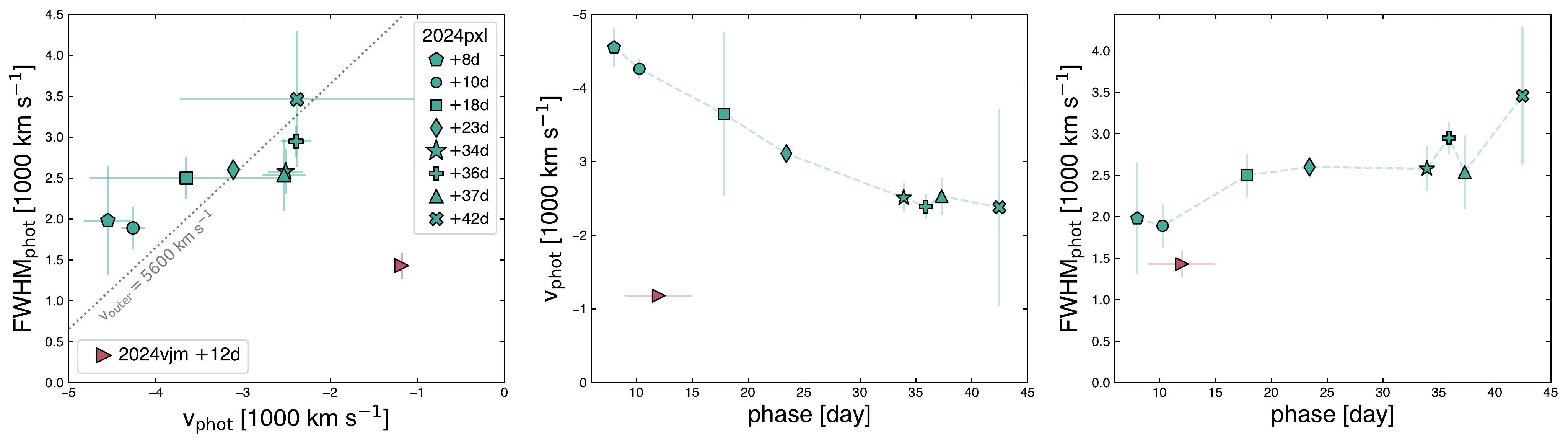}
    \caption{Photospheric NIR \ion{Co}{2} velocity and width measurements for the absorption component of the P-Cygni profile and their evolution. The dotted line represents an outer velocity edge of the ejecta where $v_\mathrm{outer}\approx v_\mathrm{phot}\;+\;$FWHM$_\mathrm{phot}$.}
    \label{fig:CoII_vels}
\end{figure*}

\subsection{IME Line Profiles}
The middle panel of \autoref{fig:phot_profiles} shows relatively isolated permitted lines of \ion{Si}{2} and \ion{Ca}{2}, which we select to represent the bulk of IMEs. The Ca and Si emission in SN~2024pxl is very weak, but similar in width to the same lines in SN~2024vjm (though the phases are different). Like the \ion{O}{1} lines, the peaks of these lines are slightly redshifted in SN~2024vjm and slightly blueshifted in SN~2024pxl. \ion{Ca}{2} and \ion{Si}{2} do not appear to have the same asymmetric profile as \ion{O}{1}, lacking the broad red shoulder.

In SN~2024vjm, \ion{Ca}{2}~2.71\um\ displays a P-Cygni profile, with photospheric absorption velocity $v_\mathrm{phot}\sim-1200$\kms. Other \ion{Ca}{2} lines are more blended, making it difficult to discern, but \ion{Ca}{2}~1.885\um\ may also show absorption. The absorption components of the P-Cygni profiles for \ion{Si}{2} in SN~2024vjm are not clearly observed, probably due to the blending of nearby weak lines. No absorption component is observed for the \ion{Ca}{2}~2.71\um\ line in SN~2024pxl, although it is unclear whether this is due to the weak line strength or if this permitted line is optically thin.

\subsection{IGE Line Profiles}
Both SN~2024pxl and SN~2024vjm exhibit rich photospheric \ion{Co}{2} line structure in the NIR. \autoref{fig:phot_profiles}, \textit{right}, displays the most isolated of these lines, \ion{Co}{2}~2.46\um. Both SN exhibit clear P-Cygni profiles, with $v_\mathrm{phot}\sim-3000$\kms\ for SN~2024pxl and $v_\mathrm{phot}\sim-1500$\kms\ for SN~2024vjm. \ion{Ni}{2}~2.85\um\ and \ion{Ni}{2}~3.45\um\ are also relatively isolated permitted IGE lines in both SN, although they do not display an obvious P-Cygni absorption component. This may be attributed to blending with nearby weaker lines, or these lines may be optically thin. We find that within each SN, the IGE element lines are consistent in width, kinematic offset, and morphology with each other and also quite similar to the properties of the LME and IME permitted lines. This seems to indicate that the ejecta distributions of these different burning products are well mixed.

\subsection{NIR Co II Velocities \label{sec:phot_vels}}

\subsubsection{Velocity measurement procedure}

As described above, the \ion{Co}{2} lines between 1.5--2.5\um\ exhibit rich spectral structure and display the clearest P-Cygni profiles. Other isolated permitted lines do not have as prominent an absorption component. Photospheric velocities are measured from the trough of this absorption component (see \citealt{Singh2025} for optical line velocities of SN~2024pxl), so here we measure photospheric velocities from the \ion{Co}{2} lines. This measurement reflects the photosphere of the Co-rich part of the ejecta. We perform this analysis on all of the NIR spectra presented by \cite{Singh2025}, except for the earliest two which do not yet display clear NIR \ion{Co}{2} lines, as well as the \textit{JWST} spectrum at $+$37~days which covers 1.7--2.5\um.

We model the P-Cygni profiles as a composite of two Gaussians, a negative one for the absorption and a positive one for the emission components of the profile. Using our N1def model line list for lines and relative strengths of all \ion{Co}{2} lines in the 1.5--2.5\um\ region, we model every \ion{Co}{2} line with this ``P-Cygni'' double-Gaussian profile and sum the contribution from each individual line to create a model for the \ion{Co}{2}. We constrain the widths and velocities of the absorption and emission Gaussians to be the same for all lines, and then fit this model to the data. 

We measure the photospheric velocity in this way to account for contributions from weaker nearby lines, which blend with the strongest lines and shift where the minima of the absorption troughs appear. Especially at wavelengths shorter than 2.0\um, there are many permitted lines with absorption components that overlap (note that \autoref{fig:24pxl_lineID} and \autoref{fig:24vjm_lineID} only label the strongest lines in this region), so the most accurate photospheric velocity measurements will be made when fitting all contributing \ion{Co}{2} lines simultaneously. For individual lines, the \ion{Co}{2}~2.46\um\ line is the most isolated, especially in SN~2024vjm, but it is only resolved in the \textit{JWST} data as it is very close to the ground-based NIR limit of $\sim$2.5\um.

The photospheric velocity is expected to vary as a function of wavelength, because the optical depth changes. Thus, fitting all \ion{Co}{2} lines across a large wavelength range may introduce uncertainties into the photospheric velocity measurement as it optimizes for a single velocity for all lines. We explore fits across several wavelength ranges within 1.5--2.5\um\ and find general consistency, with any change in photospheric velocity over wavelength being relatively small. We therefore choose to measure \ion{Co}{2} velocities between 1.5-1.85\um\ because it covers two clusters of strong \ion{Co}{2} lines and has higher S/N in the ground-based spectra. All fits are done on the unbinned spectra. To conservatively determine uncertainties, we generate 100 perturbed (noise added) realizations of the original spectrum (adopting S/N $\approx$ 5) and take the standard deviation of the fit parameters as our uncertainty estimate.

\subsubsection{Co II velocities and FWHM}

\autoref{fig:CoII_vels} shows our \ion{Co}{2} velocity measurements for SN~2024pxl and SN~2024vjm, and values can be found in \autoref{tab:coii_vels}.  As expected, SN~2024vjm displays narrower line widths and lower photospheric velocities than SN~2024pxl, at all epochs. In SN~2024pxl, the photospheric velocities decrease over time as the ejecta expands and dilutes and the spectrum-forming region recedes to more interior layers. In the optical, the \ion{Si}{2} velocities drop to $\sim$2400\kms\ by $+$5~days, whereas the \ion{Co}{2} velocities are much higher, at $\sim$4000\kms\ at $+$10~days \citep{Singh2025}. This reflects the composition inversion and mixing expected in pure deflagrations \citep{Phillips2007}. Our velocity measurements are roughly consistent with those of \cite{Singh2025}, measured by directly fitting the absorption trough, rather than fitting all lines to account for line blending. \cite{Singh2025} show that the \ion{Fe}{2} and \ion{Co}{2} photospheric velocities are higher than those of \ion{Si}{2}. \cite{Stritzinger2015} and \cite{Tomasella2016} find relatively good agreement between optical Fe and NIR Co velocity measurements. 

\begin{figure*}
    \centering
    \includegraphics[width=\linewidth]{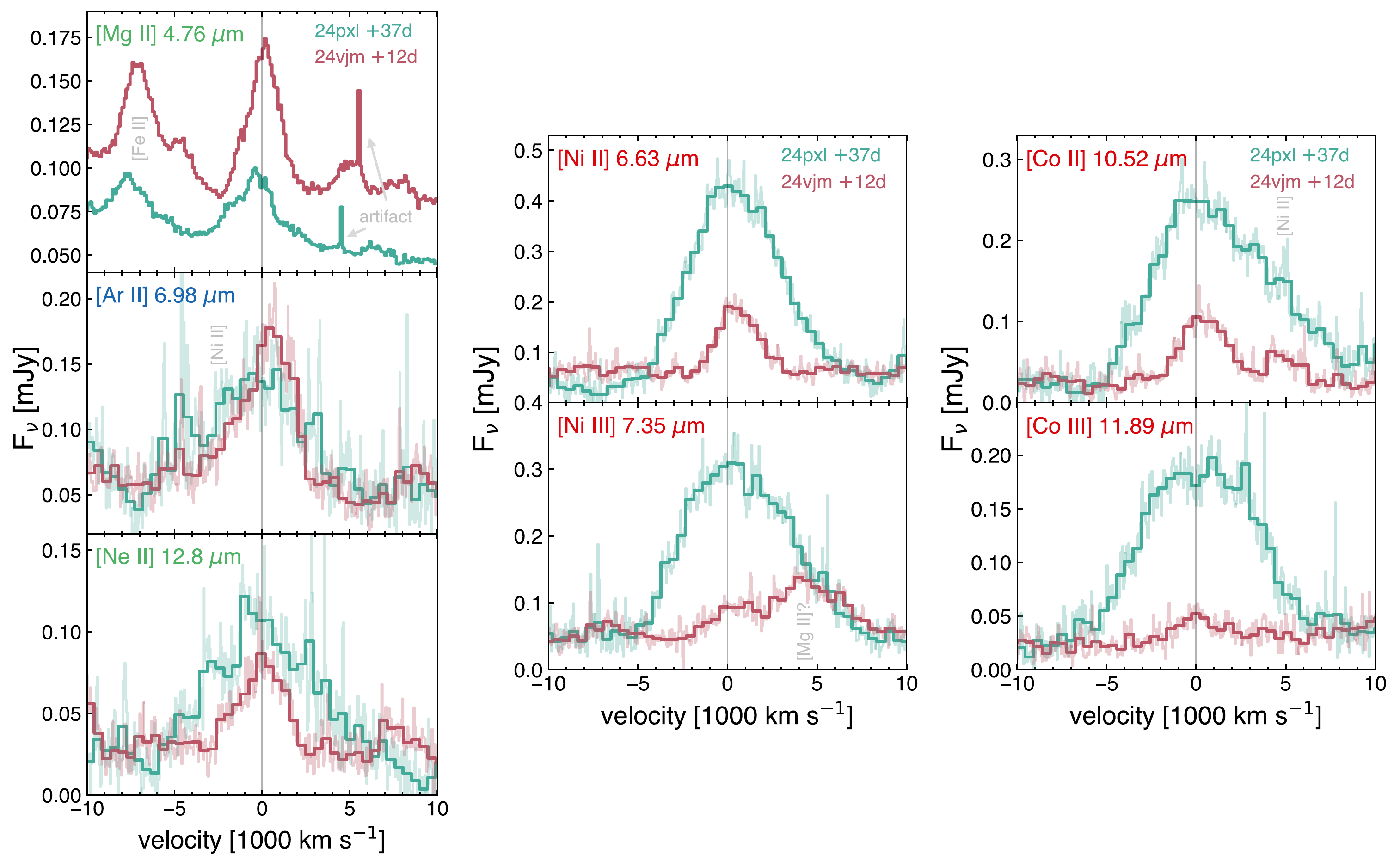}
    \caption{Comparison of selected relatively isolated forbidden emission lines in SN~2024pxl at $+$37~days (teal) and SN~2024vjm (pink) at $+$12~days. The low opacity lines show the unbinned data. The spectra have not been scaled or offset.}
    \label{fig:neb_profiles}
\end{figure*}

The FWHM of the absorption component shows an increasing trend with time. This may indicate the growth of the \ion{Co}{2} line-forming region, assuming the greater part of the ejecta has similar Co abundances and constant/slowly decreasing temperature, which may be an effect of ionization changes. As photospheric velocity ($v_\mathrm{phot}$) declines and width (FWHM) increases, the sum of these velocities ($v_\mathrm{outer}$) stays roughly constant, between $\sim$5000--6000~\kms. The left panel of \autoref{fig:CoII_vels} shows this trend, where the line showing a constant $v_\mathrm{outer}=$5600\kms\ loosely follows the data. This may imply that there is an outer edge ($v_\mathrm{outer}$) in the emitting region representing a drop in density. If this is the case, all lines should be affected; however, in practice this is difficult to measure for other lines which are more blended or do not display a clear P-Cygni profile. More data should be collected at earlier and later phases and in other SN~Iax to confirm this trend.

\section{Forbidden Lines \label{sec:neb_line_profiles}}

At the early phase of $+$11~days, we observe forbidden emission at wavelengths $\lambda\gtrsim$ 3\um\ in both SN~2024pxl and SN~2024vjm. Other low-luminosity SN~Iax have exhibited forbidden emission in their optical spectra at early times $\sim+$30~days \citep{Singh2023}, but they never go fully nebular \citep{Jha2006}. Permitted lines displaying a P-Cygni profile still dominate the optical spectra for $>$500~days \citep{Camacho-Neves2023}. In our \textit{JWST} observations, we see permitted and forbidden lines intermingling between 3--6\um, but we do not observe permitted lines at $\lambda>$ 6\um. A continuum is observed beneath the emerging MIR forbidden lines at longer wavelengths. This continuum fades over time in SN~2024pxl, which, in conjunction with the increasing line strength of the MIR forbidden lines, we interpret as the ejecta transitioning to increasingly optically thin at MIR wavelengths. This MIR transition to being dominated by forbidden emission lines occurs at much earlier times than in the optical for normal SN~Ia. In \autoref{fig:neb_profiles}, we compare the line profiles of SN~2024pxl at $+$37~days and SN~2024vjm at $+$12~days, in velocity space, of several important forbidden lines.

\begin{figure*}
    \centering
    \includegraphics[width=\linewidth]{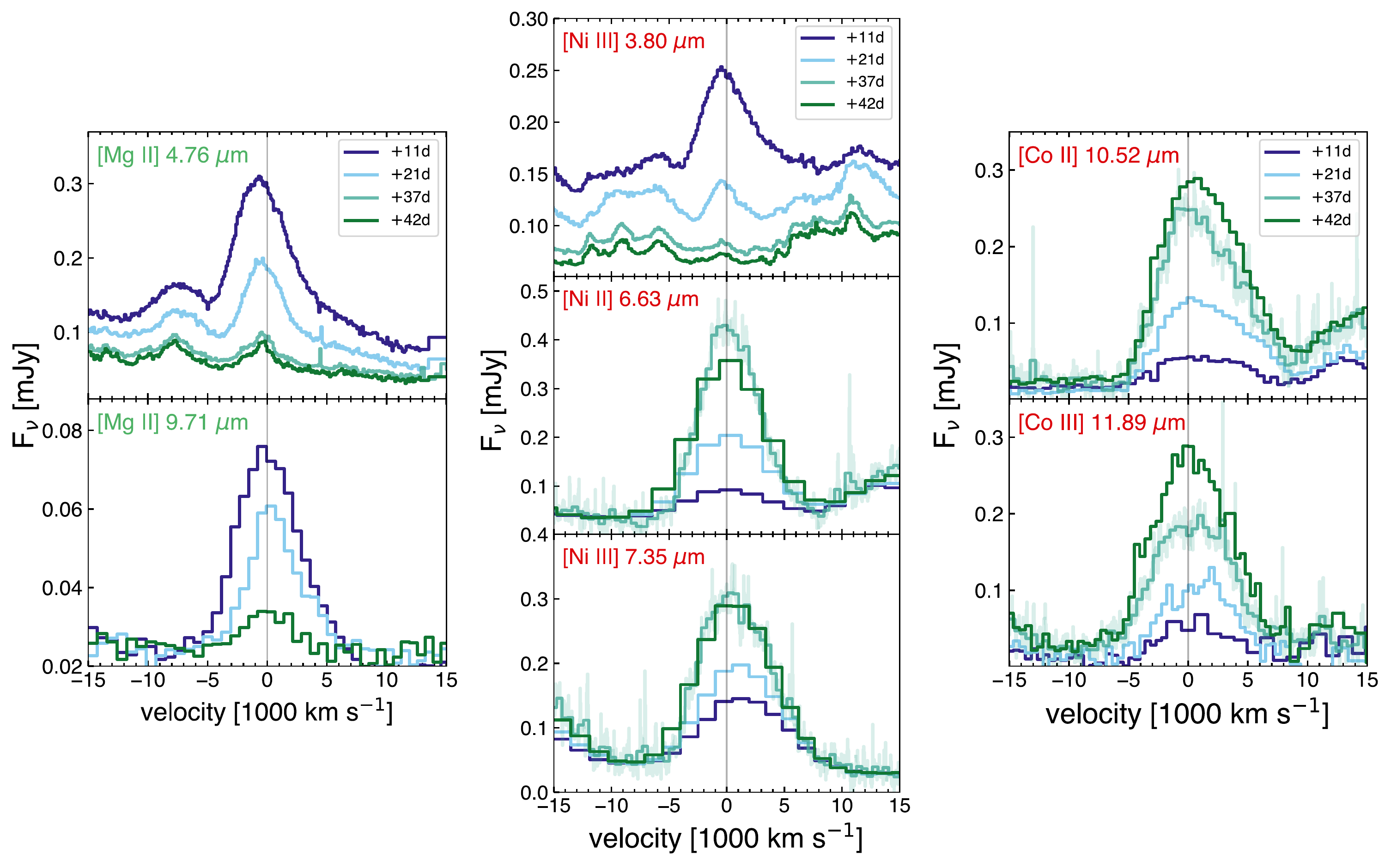}
    \caption{Evolution of Mg, Ni, and Co forbidden emission line profiles of SN~2024pxl at $+$11, $+$21, $+$37, and $+$42~days. For lines at wavelengths $>$5\um, spectra from the different epochs are offset for clearer comparison.}
    \label{fig:line_evolve}
\end{figure*}

\subsection{LME and IME Line Profiles}

The most prominent and isolated low-mass element (LME; C, O, Ne, Mg) and IME forbidden lines in SN~2024pxl and SN~2024vjm are [\ion{Mg}{2}]~4.76\um, [\ion{Ar}{2}]~6.98\um, and [\ion{Ne}{2}]~12.8\um, shown in \autoref{fig:neb_profiles}, \textit{left}. The overall ionization states of the SN~Iax ejecta are lower than observed in SN~2021aefx, SN~2022xkq and SN~2022pul, so the [\ion{Ar}{3}]~8.99\um\ and [\ion{Ca}{4}]~3.26\um\ lines which were strong in these objects are very weak or not detected in the SN~Iax. In this work, we group C, O, Ne, and Mg as LMEs because O, Ne, and Mg are products of carbon burning, and/or could be unburned material in the case of an O/Ne or O/Ne/Mg WD progenitor. 

The line velocity profiles of SN~2024vjm show remarkable agreement in morphology between Mg, Ar, and Ne, each exhibiting a centrally peaked but slightly asymmetric morphology that leans toward the redshifted side. This morphology is also roughly consistent with that of permitted \ion{O}{1}. In SN~2024pxl, these same line profiles have a more blueshifted peak compared to SN~2024vjm, and are somewhat broader. The MRS data are somewhat noisy and there may be more contamination from line overlap in SN~2024pxl; however, it appears that Mg, Ar, and Ne are not as smoothly peaked and may show signs of clumping in the ejecta. Interestingly, the [\ion{Mg}{2}]~4.76\um\ line profile in SN~2024pxl is nearly flipped compared to the \ion{O}{1} profiles, with a peak on the redder side and a bluer shoulder. The absence of a narrow feature at the same velocity as in \ion{O}{1} is consistent with the predictions of \cite{Fink2014} of an unburned O channel and suggests that most of the Mg in SN~2024pxl is likely a product of C burning.

In both SN, the profiles exhibit slight asymmetries deviating from a smooth Gaussian, suggesting variations from a spherical geometry which might arise from bulges, bubbles, or clumps in the ejecta. Most importantly, the central peak of these profiles indicates that the LMEs and IMEs must be located in the central regions of the ejecta. This is in direct contrast to the flat-topped morphologies observed in SN~2021aefx and SN~2022pxl at late-times (i.e., the lines are optically thin) for Ar and Ca lines, implying that the SN~Iax ejecta is chemically well-mixed rather than stratified. As we discuss further in \autoref{sec:conclusions}, a mixed ejecta structure is a signature of a deflagration.

\subsection{IGE Line Profiles}

In SN~2024pxl, the IGEs have a smoother and more symmetric distribution than the IMEs (see \autoref{fig:neb_profiles}, \textit{middle} and \textit{right}, but note that the IGE lines are stronger and have higher S/N, which may contribute to them appearing smoother). The [\ion{Ni}{2}]~6.64\um\ and [\ion{Co}{2}]~10.52\um\ line profiles (taking into account the [\ion{Ni}{2}]~10.67\um\ line on the red side of [\ion{Co}{2}]) are marginally thinner and blueshifted compared to the higher ionization [\ion{Ni}{3}]~7.35\um\ and [\ion{Co}{3}]~11.89\um\ line profiles. Narrower widths are expected for lower ionization states of IGEs: the densest inner regions of the ejecta make recombination more favorable, producing lower ionization states. The slight blueshift of these singly ionized lines relative to the doubly ionized lines may hint at a small offset in the ejecta distributions.
Alternatively, the inner regions with a lower ionization may not be fully optically thin yet, and some of the redshifted side of the ejecta might still be obscured. In SN~2024vjm, the IGE profiles are somewhat narrower than the LME and IME profiles.

\begin{figure*}
    \centering
    \includegraphics[width=\linewidth]{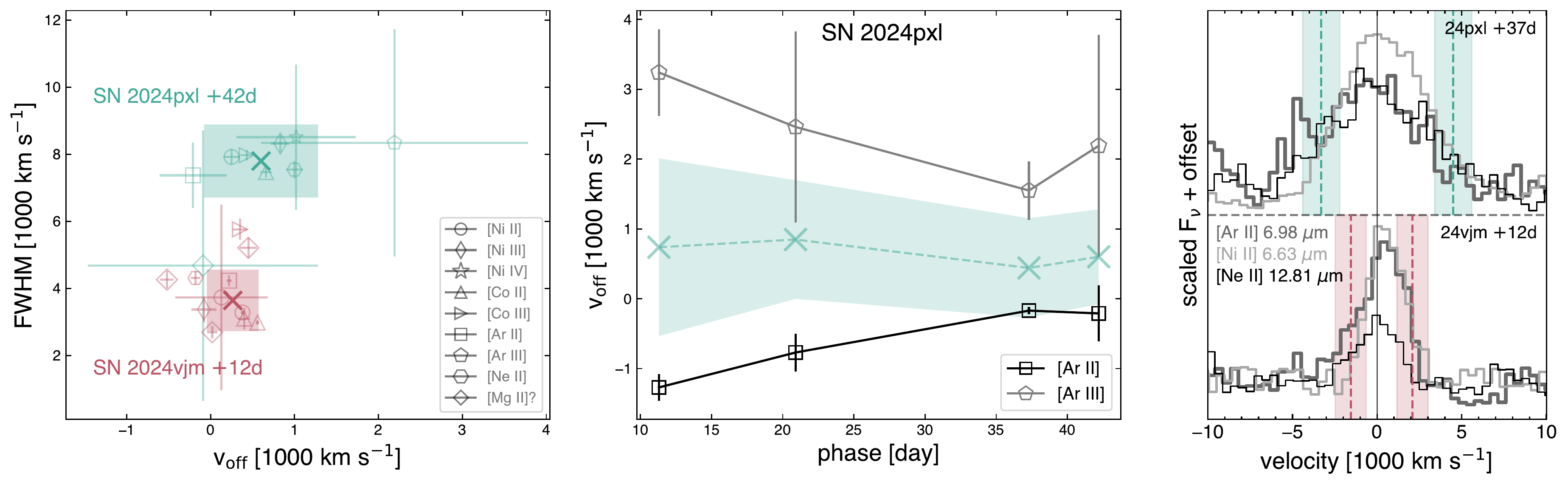}
    \caption{\textit{Left}: FWHM vs. kinematic offset, v$_\mathrm{off}$, of MIR forbidden emission lines of SN~2024pxl at $+$42~days and SN~2024vjm at $+$12~days. Heavy crosses represent the uncertainty weighted average of all lines, and shaded regions represent the standard deviation. \textit{Middle}: Evolution of the bulk motion in SN~2024pxl (teal), which remains constant, and the Ar lines, which show an ionization offset. \textit{Right}: Forbidden line profiles of [\ion{Ar}{2}] (medium gray, thick line width), [\ion{Ni}{2}] (light gray), and [\ion{Ne}{2}] (black, thin line width) with color panels showing the average FWHM and standard deviation. The lines have similar profile shapes and widths, indicating the ejecta are well-mixed.}
    \label{fig:24pxl_neb_vels}
\end{figure*}

Although MIR IGE emission in SN~2024pxl is relatively strong compared to LME and IME emission, the relative strength of IGE emission to LME and IME emission in SN~2024vjm is significantly lower. The middle and right panels of \autoref{fig:neb_profiles} show that in SN~2024vjm, [\ion{Ni}{2}]~6.63\um\ and [\ion{Co}{2}]~10.52\um\ (which is more isolated in SN~2024vjm due to the lower velocities) are consistent in morphology with the LMEs and IMEs, and that the ionization is lower than in SN~2024pxl. In fact, [\ion{Co}{3}]~11.89\um\ is only weakly detected in SN~2024vjm, and the [\ion{Ni}{3}]~7.35\um\ line is only half the strength of a new neighboring line at 7.45\um\ which we tentatively identify as [\ion{Mg}{2}]~7.45\um. This [\ion{Mg}{2}]~7.45\um\ line is not present, or does not contribute significantly, in SN~2024pxl. It is difficult to say from the observations alone whether this difference in relative strength of the IGE emission in SN~2024vjm is due to a smaller mass fraction of IGEs versus LMEs and IMEs produced in the explosion, or whether it is fully or partially due to the difference in phase of the observations.

\subsection{Line-Profile Evolution in SN~2024pxl}

\autoref{fig:line_evolve} shows the evolution of several quite isolated lines in SN~2024pxl which change significantly. The earliest spectrum of SN~2024pxl exhibits prominent lines of [\ion{Mg}{2}]~4.76\um\ and [\ion{Mg}{2}]~9.71\um. The left panel of \autoref{fig:line_evolve} displays how these lines decrease in strength, width, and kinematic offset over time. [\ion{Mg}{2}]~9.71\um\ is not detected above the noise in the $+$37~day MRS spectrum, so we omit this epoch, but it is still weakly detected in the higher S/N LRS spectrum at $+$42~days. These two [\ion{Mg}{2}] lines appear to evolve consistently with each other, although [\ion{Mg}{2}]~9.71\um\ appears slightly narrower with a lower peak kinematic offset in the second epoch. These fading lines are consistent with dropping densities and temperatures resulting in decreasing recombination rates.

[\ion{Ni}{3}]~3.80\um\ evolves similarly to the [\ion{Mg}{2}] lines, fading, narrowing, and slowing. However, unlike the [\ion{Mg}{2}] lines which result from recombination, this [\ion{Ni}{3}] line is collisionally excited. One of the more prominent lines in the 3--5\um\ region in the first epoch, this line is only a small bump by the last epoch. As mentioned in \autoref{sec:lineIDs}, these disappearing forbidden IGE lines arise from transitions between excited states. In comparison, the MIR IGE emission lines arising from low-lying transitions to the ground state (which can be collisionally excited) such as [\ion{Ni}{2}]~6.63\um, [\ion{Ni}{3}]~7.35\um\ grow in strength (see \autoref{fig:line_evolve}, \textit{middle} and \textit{right}). Non-thermal excitation is subdominant here, so these higher energy level transitions decrease over time with the ejecta densities and temperatures. Instead, lower-energy transitions to the ground state begin to dominate. We note, however, that this effect on the observed fading forbidden IGE lines may be degenerate with a decrease in ionization state.

The [\ion{Co}{2}]~10.52\um\ and [\ion{Co}{3}]~11.89\um\ emission are fairly isolated (the red side of [\ion{Co}{2}]~10.52\um\ is slightly contaminated by a weaker [\ion{Ni}{2}]~10.68\um\ line) in SN~2024pxl, as well as decently well resolved. \autoref{fig:line_evolve} shows the emergence of these strong MIR forbidden Co lines over time. [\ion{Co}{2}]~10.52\um\ at $+12$~days is relatively flat above the continuum with FWHM $\sim$ 8,500\kms, indicating that Co is not yet fully optically thin and the line-forming region is at $v\gtrsim$8,500\kms. By $+22$~days, we are seeing deeper into the blueshifted side of the ejecta, while the slanted slope for $v>0$\kms\ indicates the redshifted side is still partially obscured. The line appears to be mostly optically thin by $+$42~days, where we see a peaked, symmetric profile. [\ion{Co}{3}]~11.89\um\ evolves similarly, although the data in this line suffer from more noise at the long-wavelength end of LRS. The resolution at short wavelengths in LRS is poor so subtle changes in the shape of the [\ion{Ni}{2}]~6.63\um\ and [\ion{Ni}{3}]~7.35\um\ lines cannot be determined, but they appear to behave consistently with the Co lines.

\subsection{Forbidden-Line Velocities \label{sec:neb_vels}}

To measure velocities from the line profiles, we first isolate the lines themselves by subtracting the lingering thermal continuum (likely from the contribution of a remnant, not from the SN ejecta) at each epoch. Following the procedure described by \cite{Kwok2024}, we use our N1def model line list (introduced in \autoref{sec:lineIDs} and further discussed in \autoref{sec:models}) to set the expected flux ratios between lines of the same ion and fit all lines between 5--14\um\ for LRS epochs and 5--20\um\ for MRS epochs. These lines are fit simultaneously, requiring that all lines from the same ion have the same FWHM and kinematic offset (v$_\mathrm{off}$). Uncertainties are estimated using the same procedure as in \autoref{sec:phot_vels}. The uncertainties in the measurements from the $+$37~day MRS spectrum are smaller due to the much higher resolution. Uncertainties in the LRS wavelength solution in addition to the low resolution contribute to large uncertainties.

\subsubsection{SN~2024pxl velocities}

\autoref{fig:24pxl_neb_vels} shows our measured values of kinematic offset (v$_\mathrm{off}$) and FWHM for SN~2024pxl at the final epoch ($+$42~days) when the ejecta is mostly optically thin (left panel). All measured values are given in \autoref{tab:24pxl_neb_vels}. Most of these measurements change only marginally over time, except for the [\ion{Mg}{2}]~9.71\um\ line, which narrows. The middle panel shows that the bulk kinematic velocity (teal), obtained by averaging all measured lines weighted by their uncertainty, stays flat at $\sim$1000\kms. The IGE velocities cluster around this bulk motion. The Ar lines, however, start offset from the bulk motion and gradually converge toward it, and display an ionization stratification where [\ion{Ar}{3}] is redshifted but [\ion{Ar}{2}] is blueshifted. This offset, especially in the first epoch, is significant; however, given the LRS wavelength uncertainty at the [\ion{Ar}{2}]~6.98\um\ line, and potential contaminating line overlap from [\ion{Mg}{2}]~6.94\um\ and [\ion{Ni}{2}]~6.92\um, it is difficult to discern if this represents a physical difference in the ejecta and what process would create it. The right panel of \autoref{fig:24pxl_neb_vels} shows that [\ion{Ar}{2}], [\ion{Ni}{2}], and [\ion{Ne}{2}], representative of the IMEs, IGEs, and LMEs, all have similar profiles and FWHM, indicating that the ejecta are well-mixed. If the ejecta were highly stratified, these measurements should instead show offsets in FWHM, such as that seen between the Ar lines versus the IGEs in SN~2021aefx \citep{Kwok2023}.

\cite{Fink2014} find that offsets in the deflagration ignition (i.e., off-center explosions) produce bound remnant velocity kicks on the order of tens of \kms, while \cite{Lach2022} find kicks velocities on the order of a few hundred \kms. These are consistent with the bulk motion of SN~2024pxl.

% \begin{figure}
%     \centering
%     \includegraphics[width=\linewidth]{24vjm_nebular_velocities.pdf}
%     \caption{FWHM vs. kinematic offset, v$_\mathrm{off}$, of MIR nebular emission lines of SN~2024vjm at $+$12~days.}
%     \label{fig:24vjm_neb_vels}
% \end{figure}

\subsubsection{SN~2024vjm velocities}

We measure velocities for the MRS spectrum of SN~2024vjm, out to 16\um\ where the data are still high enough S/N, in a similar way as described above for SN~2024pxl. However, as discussed in more detail in \autoref{sec:models}, the N1def model from \cite{Fink2014} differs significantly in terms of predicted $^{56}$Ni and ejecta masses compared to observations of extremely low-luminosity SN~Iax such as SN~2024vjm. Thus, we fit individual lines separately, rather than enforcing predicted line ratio strengths that are calculated based on the density and temperature of the N1def model, which is likely significantly different than that of SN~2024vjm. We do fit [\ion{Mg}{2}]~7.45\um\ together with [\ion{Ni}{3}]~7.35\um, and [\ion{Co}{2}]~10.52\um\ with [\ion{Ni}{2}]~10.67\um, as these lines are close to each other and we want to account for line blending. All other lines in SN~2024vjm are isolated enough to fit alone. The values of these measurements can be found in \autoref{tab:24vjm_neb_vels}.

By taking the uncertainty-weighted average of the parameters of all measured lines, we find SN~2024vjm has a bulk kinematic offset of $\sim$250\kms\ and narrower lines with FWHM $\sim$ 3600\kms. Similar to SN~2024pxl, the right panel of \autoref{fig:24pxl_neb_vels} shows that the IMEs, IGEs, and LMEs display similar line profiles and widths consistent with well-mixed ejecta. No clear signs of separation or stratification are detected.

\begin{figure}
    \centering
    \includegraphics[width=\linewidth]{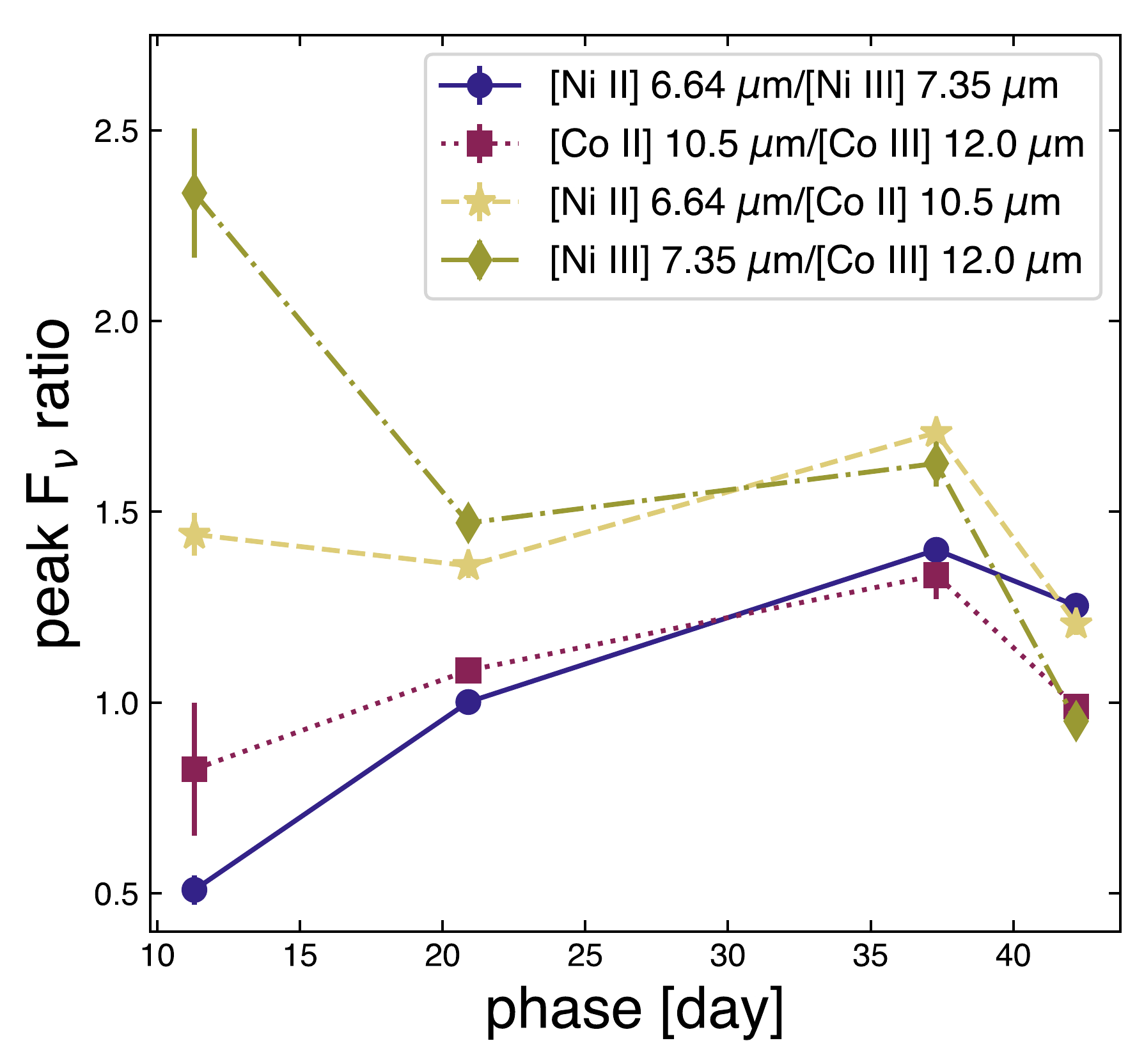}
    \caption{Evolution of the ratios between the flux density $F_\nu$ at peak of the most prominent MIR Ni and Co lines in SN~2024pxl. The continuum was subtracted before measuring the value of the peaks.}
    \label{fig:24pxl_ratios}
\end{figure}

\subsection{Line-Strength Evolution in SN~2024pxl \label{sec:ion_ratios}}

We also measure the flux density ($F_\nu$) at the peak of the prominent MIR forbidden lines in SN~2024pxl at each epoch, after continuum subtraction. We measure peak fluxes, rather than integrating line fluxes, because we can measure these directly, while line overlap can contaminate widths (especially in the [\ion{Co}{2}]~10.52\um\ line which blends with [\ion{Ni}{2}]~10.68\um). These measurements are given in \autoref{tab:peak_flux}, and uncertainties were calculated with a bootstrapping method, similar to that described in \autoref{sec:phot_vels}. At these early epochs, the change in forbidden line strength may be attributed to a complex combination of factors, including a change in composition from the decay of $^{56}$Ni to $^{56}$Co; the contribution of flux from ejecta regions which were previously optically thick but become optically thin; and shifts in ionization such that flux increases in some lines and decreases in others. Fully disentangling the individual effects of these factors will likely require the aid of models. Nonetheless, we identify two trends in the evolving line strengths with physical implications.

The ratio of [\ion{Ni}{2}]/[\ion{Ni}
{3}] and [\ion{Co}{2}]/[\ion{Co}{3}] reflects the changing ionization of the ejecta, since the ratio is between different ionization states of the same element. \autoref{fig:24pxl_ratios} shows that Ni and Co follow a similar ionization trend where the ionization state decreases over the first three epochs and then begins to fall off again in the final epoch. This may reflect an increase in flux from more central regions that are becoming optically thin, as the ionization is lower in the center due to higher densities that can facilitate recombination. However, the densities will continue to drop and eventually recombination to [\ion{Ni}{2}] becomes less favorable.

Comparison along the same ionization state between different elements with the [\ion{Ni}{2}]/[\ion{Co}{2}] and [\ion{Ni}{3}]/[\ion{Co}{3}] ratios can probe a change in composition in the emitting region. These ratios exhibit similar trends (see \autoref{fig:24pxl_ratios}). Radioactive decay should create an exponential decay; we see a general decreasing trend, but it is not exponential. Our trend suggests that emission from Ni drops, increases, then drops again, relative to Co.

\begin{figure*}
    \centering
    \includegraphics[width=\linewidth]{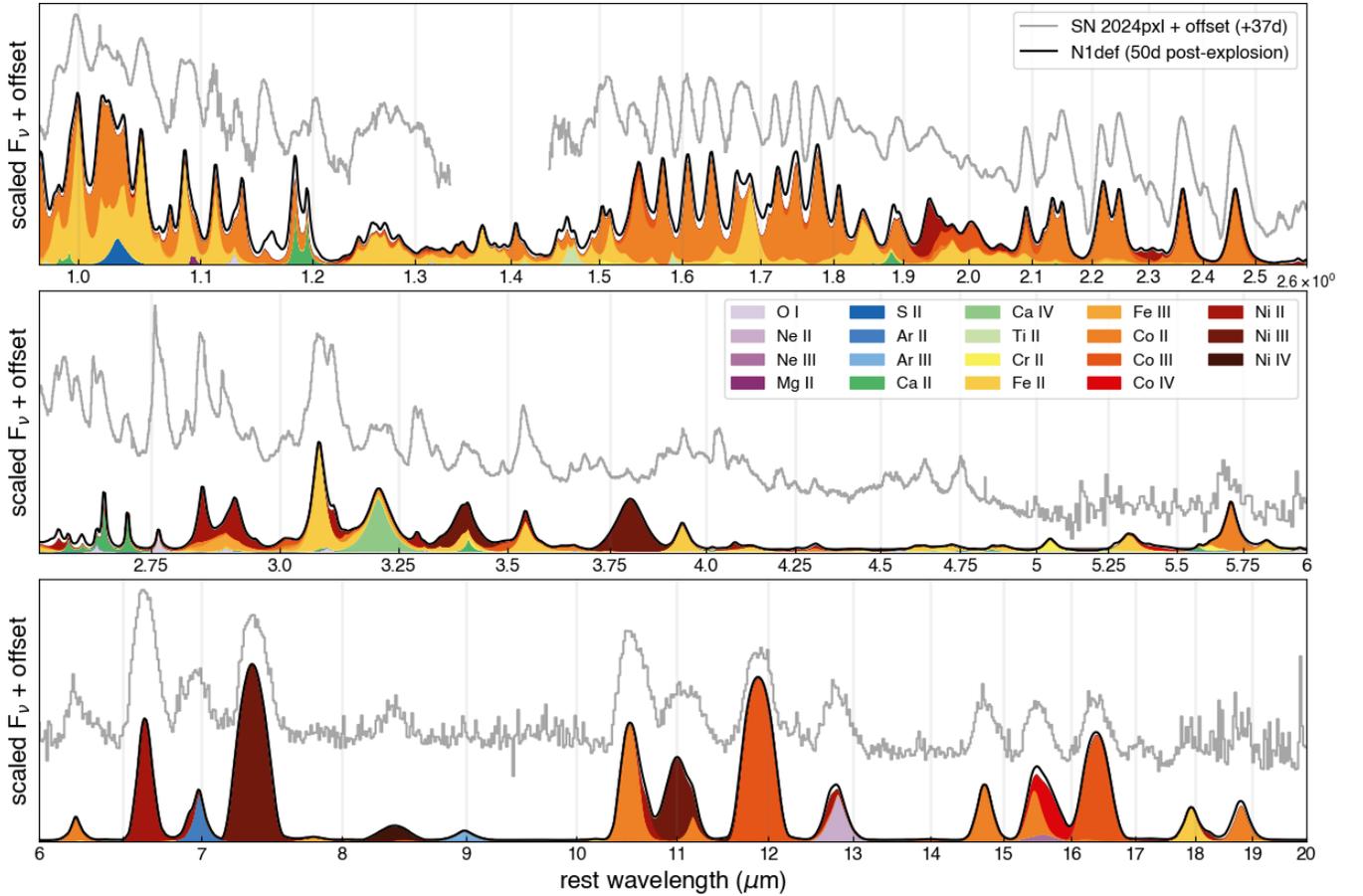}
    \caption{N1def model spectrum at 50~days post-explosion (black) compared with SN~2024pxl at $+$37~days, corresponding to 50~days post-explosion (gray). The SN~2024pxl spectrum has been offset for clarity, and the flux is displayed using an arcsinh scaling to show details of weaker lines. Contributions from each ion to spectral features are shaded by color.}
    \label{fig:n1def_SDEC}
\end{figure*}

\section{Models \label{sec:models}}

High-luminosity SN~Iax such as SN~2012Z \citep{Stritzinger2015}, SN~2005hk \citep{Phillips2007}, and SN~2020udy \citep{Maguire2023, Singh2024},  as well as intermediate-luminosity objects such as SN~2002cx \citep{Li2003}, SN~2014ck \citep{Tomasella2016}, and SN~2019muj \citep{Barna2021b}, are generally consistent with existing deflagration models \citep{Fink2014, Lach2022} and constraints on companion stars \citep{McCully2014, Foley2014b, Maguire2023}. \cite{Fink2014} parameterize the strength of their deflagration models by the number of ignition points. More ignition points produce a higher explosion energy, resulting in larger ejected mass and higher $^{56}$Ni production that directly creates higher luminosities. \cite{Lach2022} also find that their variations on a single-ignition-point explosion form a one-dimensional sequence where the main characteristic is the mass of $^{56}$Ni. Synthetic spectra calculated through the radiative transfer code TARDIS, using density and abundance profiles from the lowest energy deflagration model (N1def, single ignition point) from \cite{Fink2014}, match observations of SN~2014dt, a relatively luminous SN~Ia, exceptionally well even out to $>$500~days post-explosion \citep{Camacho-Neves2023}. 

However, it is difficult to explain the lowest luminosity SN~Iax such as SN~2024vjm. Several studies have attempted to model low-luminosity SN~Iax \citep{Kromer2015, Kashyap2018, Lach2022}, but none have been able to reproduce all their properties. In particular, while some can roughly reproduce the small amounts of synthesized $^{56}$Ni needed to match the peak luminosities of low-luminosity SN~Iax light curves, the resulting transients evolve on a much faster timescale than is actually observed in low-luminosity events such as SN~2008ha, SN~2010ae, SN~2020kyg, and SN~2021fcg. The timescale is related to the ejecta mass, so the models have lower ejecta masses than calculated from the observations. For example, the N5def\_hybrid model from \cite{Kromer2015} produces roughly the right $^{56}$Ni mass to match the observed luminosities ($\sim$\num{3e-3}\msun); however, it does not produce enough total ejecta mass (0.014\msun\ vs. 0.15\msun), causing the model transient to fade far more quickly than the observations. 

Alternatively, the lowest luminosity SN~Iax may arise from entirely different mechanisms. \cite{Karambelkar2021} suggest that SN~2021fcg, the faintest observed SN~Iax to-date at $M_r\sim -12.7$~mag, may have originated from the merger of a carbon-oxygen (C/O) WD and an oxygen-neon (O/Ne) WD such as that described by \cite{Kashyap2018}, which is predicted to create a very low-luminosity transient. It remains to be seen whether such transients would spectroscopically resemble low-luminosity SN~Iax.

\begin{figure*}
    \centering
    \includegraphics[width=\linewidth]{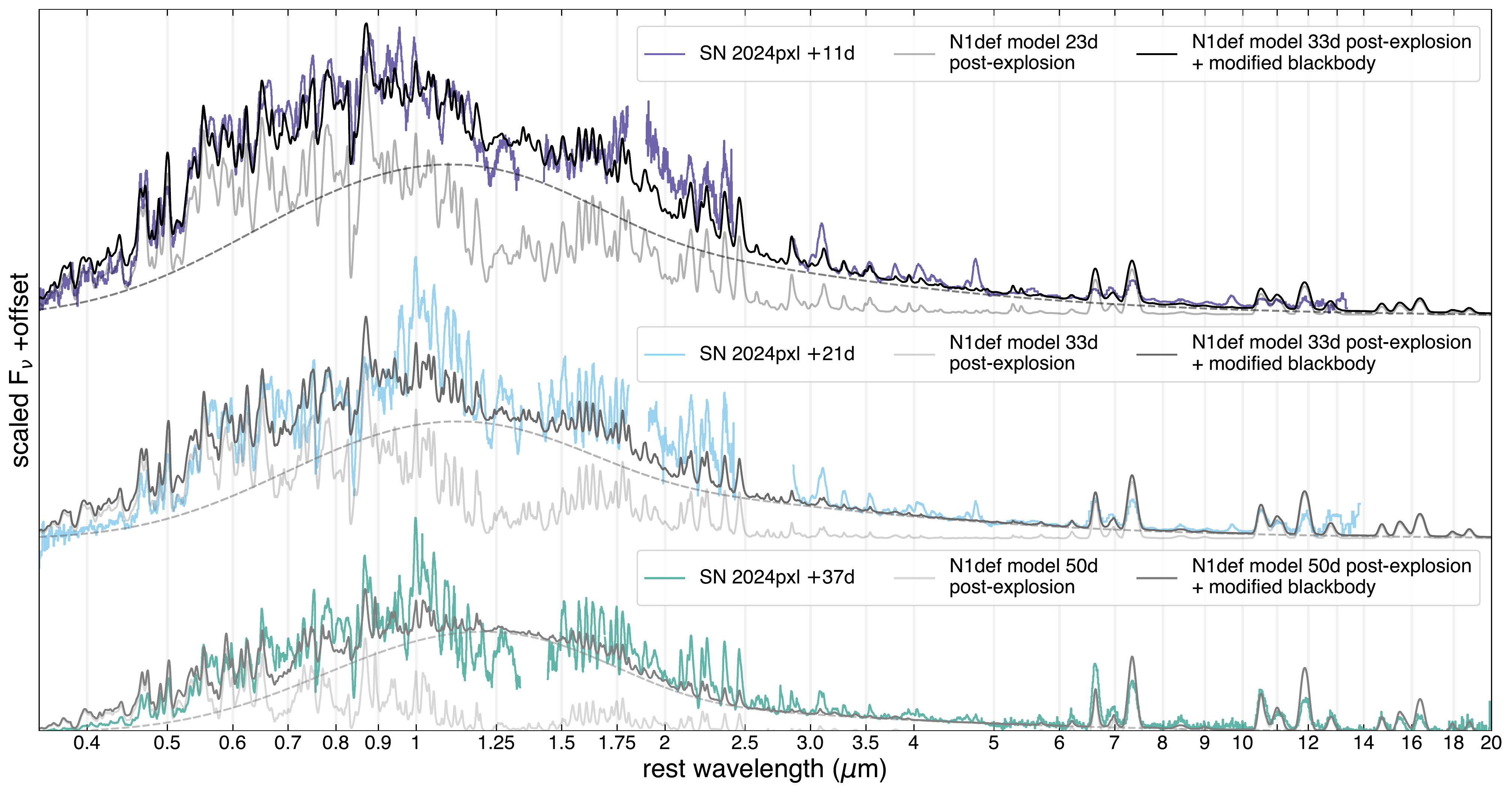}
    \caption{SN~2024pxl at $+$11~days (23~days post-explosion; indigo), $+$21~days (33~days post-explosion; light blue) and $+$37~days (50~days post-explosion; teal) compared to the N1def model at 23, 33 and 50~days post-explosion (gray, light gray, and lightest gray, respectively). Adding illustrative modified $\sim$4000~K blackbodies to the model spectra (black, dark gray, and gray, respectively) improves the match of the continuum flux, potentially indicating needed effects from time-dependent models or additional energy input from a remnant. The modified blackbodies are added post radiative transfer and are therefore not physically accurate. The flux is displayed using an arcsinh scaling to show details of weaker lines. Models are scaled to a distance of 23~Mpc \citep{Singh2025}.}
    \label{fig:24pxl_n1def}
\end{figure*}

\subsection{Radiative Transfer Models}

Our radiative transfer modeling follows a similar procedure as outlined by \cite{Blondin2023}. Starting with the spherically-averaged density and abundance profiles at $t\approx 100$\,s post explosion available on HESMA, we generated initial conditions at a later time (23\,d and 33\,d post explosion for the N1def model, 20\,d post explosion for the N5def\_hybrid model; see following sections) taking into account changes in composition induced by the decay of radioactive isotopes and the decrease in density due to homologous expansion ($\rho \propto 1/t^3$). We applied a small radial mixing to the HESMA inputs with a characteristic velocity width $\Delta v_{\rm mix}=200$\,\kms\ to smooth sharp variations in composition \citep[see, e.g.,][]{Blondin2022b}. We then solve the 1D non-LTE radiative transfer with CMFGEN \citep{Hillier2012} assuming steady state. Given the predicted small ejecta masses for lower-luminosity SN~Iax, which turn optically thin more quickly, steady-state models at these phases are still reasonable and instructive, especially in the IR. Time-dependent effects become increasingly important at earlier phases. However, future modeling of these systems should include time-dependent effects, which are beyond the scope of the current study. Furthermore, our modeling approach ignores the contribution of the bound remnant predicted by both models, which has been shown to affect the radiative display \citep[e.g.,][]{Callan2024}.

Non-local energy deposition from radioactive decay was treated using a Monte-Carlo approach for $\gamma$-ray transport. Non-thermal processes are accounted for through a solution of the Spencer-Fano equation \citep[see][]{Li2012}. The following ions were included: He\one--{\sc ii}, C\one--{\sc iii}, N\one--{\sc iii}, O\one--{\sc iii}, Ne\one--{\sc iii}, Na\one, Mg\two--{\sc iii}, Al\two--{\sc iii}, Si\two--{\sc iv}, S\two--{\sc iv}, Ar\one--{\sc iii}, Ca\two--{\sc iv}, Sc\two--{\sc iii}, Ti\two--{\sc iii}, Cr\two--{\sc iv}, Mn\two--{\sc iii}, Fe\one--{\sc v}, Co\one--{\sc iv}, and Ni\one--{\sc v}. For all of the aforementioned ions, we also consider ionizations and recombinations from the ground state of the next ionization stage (e.g. \ion{Co}{5} in the case of Co). More details concerning the atomic data can be found in Appendix~B of \cite{Blondin2023}.

\subsection{N1def Model}

We compare SN~2024pxl to CMFGEN radiative transfer calculations of the lowest-energy model, N1def, from \cite{Fink2014} (publicly available on \href{https://hesma.h-its.org/}{HESMA}, \citealt{Kromer2017}). \cite{Singh2025} find good agreement from the light curve properties of SN~2024pxl with the N1def model, so we choose this model to investigate. We also explored the r45\_d6.0\_Z model from \cite{Lach2022}, as it has similar luminosity to SN~2024pxl and a slower light curve decline than the N1def model; however, we found the model spectra to be nearly identical to the N1def model.  \autoref{fig:n1def_SDEC} shows our N1def model spectrum at 50~days post-explosion compared to the observed spectrum of SN~2024pxl at $+$37~days, corresponding to $\sim$50~days post-explosion \citep[assuming a 12~day rise-time,][]{Singh2025}. Note that many caveats regarding time-dependent effects apply at increasingly earlier phase, as the steady-state assumption becomes increasingly less valid. The N1def model produces spectral lines with similar velocities to SN~2024pxl and predicts most of the strong spectral features. The model spectrum is shaded by the contribution of each ion to particular spectral features to elucidate the amount of line blending. The model notably underpredicts the emission strength of \ion{O}{1} and [\ion{Mg}{2}] lines. Given that this N1def model has not been tailored to SN~2024pxl in any way, we find that the match in the predicted lines from optical to MIR is very good.

The overall ionization of our N1def model is higher than that of the observations, seen most clearly in the weaker model flux in the \ion{Fe}{2} near 1\um, the \ion{Co}{2} between 1.5--2.5\um, and the [\ion{Ni}{2}]~6.63\um\ line. [\ion{Ni}{3}]~7.35\um\ and [\ion{Co}{3}]~11.89 and 16.4\um\ are also stronger in the model than observed in the data. This indicates that the density of the ejecta stays higher than the model predicts, since singly ionized states are more probable in denser environments that facilitate recombination, all other things being equal. \cite{Shingles2022} found that CMFGEN (and also ARTIS) radiative transfer models tend to produce overionized ejecta. Rectifying this modeling challenge may also mitigate the ionization discrepancy we find between the model and the data.

Furthermore, the thermal continuum in the data diverges significantly from that of the model, which has essentially no IR continuum (see \autoref{fig:24pxl_n1def}). We attempt to artificially ``correct'' the continuum by adding a 4000~K blackbody modified by a skewed Gaussian factor that suppresses optical and enhances NIR flux, potentially justifiable as some of the optical blackbody flux being trapped and redistributed to the IR. A physical interpretation of this modified blackbody addition to the model could be energy input from the remnant, which is not included in our radiative transfer calculations. We stress that this skewed blackbody addition is illustrative--not physically accurate--added after the radiative transfer simulation, and choices for temperature and skew factor were selected by eye. Including an additional energy source within the radiative transfer simulation will impact the calculations and change the ionization balance and spectral features, changing more than just the continuum flux. 

Although inclusion of the effects of a remnant is beyond the scope of the models in this present work, our results suggest that future modeling of SN~Iax, especially in the IR, must account for the remnant (as in \citealt{Callan2024}). Some of the discrepancies in the SED may also be mitigated in time-dependent models. Properly accounting for these effects might improve the ionization discrepancy as well.

\begin{figure*}
    \centering
    \includegraphics[width=\linewidth]{n5hybrid_model_20d_24vjm.pdf}
    \caption{SN~2024vjm at $+$12~days ($\sim$20~days post-explosion; pink) compared to the N5def\_hybrid C/O/Ne WD model (gray) and the N5def\_hybrid C/O/Ne WD model $+$ an illustrative 4500~K blackbody (black) at 20~days post-explosion. The additional blackbody improves the match to the NIR $+$ MIR continuum flux, but is added post radiative transfer and is therefore not physically accurate. The model overpredicts the optical flux at $<$1\um. The model has been scaled to a distance of 7~Mpc.}
    \label{fig:24vjm_n5hybrid_n1def}
\end{figure*}

\subsection{N5def\_hybrid Model}

We calculate a radiative transfer model for the N5def\_hybrid model from \cite{Kromer2015} (publicly available on HESMA) of the deflagration of a hybrid C/O/Ne WD at 20~days post-explosion (similar phase to the $+12$~day observation of SN~2024vjm, assuming an 8~day rise). As seen in \autoref{fig:24vjm_n5hybrid_n1def}, the model reproduces many of the observed lines. In the MIR, the model forbidden lines are wider than observed, indicating the model velocities are too high compared to SN~2024vjm. The model continuum flux is underestimated in the NIR and MIR, and overestimated in the optical. Flux in the MIR lines is significantly stronger than observed, but similar to the observations (excluding the continuum) in the NIR. Excess optical flux in the model is likely attributable to the higher luminosity of the N5def\_hybrid model as compared to SN~2024vjm. Reducing the deposited energy might improve the match in the optical, but could also result in underionization of the ejecta and diminished flux in the NIR and MIR as well. The observed significant IR continuum, not present in the model, suggests contribution from some additional energy source such as a bound remnant.

The N5def\_hybrid model is overluminous and the light curve fades too quickly compared to low-luminosity SN~Iax such as SN~2024vjm. This is also seen in radiative transfer modeling, where the constrained density profiles exceed those of the corresponding deflagration models in the outer regions \citep{Magee2016, Barna2021b, Magee2022}. Nonetheless, we find that the general spectral similarities between our N5def\_hybrid model and the observed SN~2024vjm spectrum provide compelling support for SN~2024vjm as a weak deflagration explosion.

Future modeling of low-luminosity SN~Iax such as SN~2024vjm should include the effects of a bound remnant. If the remnant can effectively keep the energy deposition high (both via thermal and radioactive decay energy), then a N5def\_hybrid$+$remnant model might still make a viable model for SN~2024vjm. Radiative transfer models that include remnant effects are required to properly compare the data of extremely low-luminosity SN~Iax, which leave behind the most massive remnants, to hydrodynamical models.

% \begin{figure*}
%     \centering
%     \includegraphics[width=\linewidth]{24vjm_model_comp.pdf}
%     \caption{Caption}
%     \label{fig:24vjm_n5hybrid}
% \end{figure*}

\section{Discussion and Conclusions \label{sec:conclusions}}

We have presented the first MIR spectra of any SN~Iax by obtaining data of the intermediate-luminosity Iax SN~2024pxl and the extremely faint Iax SN~2024vjm with \textit{JWST}. Despite their large difference in luminosity, their early panchromatic spectra show many similarities, particularly between the $+12$~days observation of SN~2024vjm and the $+$37~day observation of SN~2024pxl, including nearly the same lines, similar panchromatic SED shape, coexisting permitted and forbidden lines between 2.5--5\um, and the early emergence of forbidden emission in the MIR. The major spectral similarities between these objects may point to a similar origin for both events. 

SN~2024vjm also displays several differences from SN~2024pxl, namely narrower lines, lower overall luminosity, more prominent permitted LME and IME lines (O, Mg, Si, Ca), several additional forbidden lines tentatively identified as [\ion{Mg}{2}], and increased similarity to SN~2024pxl at a later rather than coeval phase. These properties likely indicate that SN~2024vjm has a steeper density profile, lower ejecta mass that becomes optically thin more rapidly, and an increased mass fraction of IMEs.

These early-time MIR spectra are complex, with contributions from optically thick and optically thin permitted lines, and emerging forbidden emission. Different ions have different critical densities, so in principle, a careful analysis with the aid of improved models may be able to place constraints on the densities and geometries for these various ion species within the ejecta.

The panchromatic spectra of SN~2024pxl show good agreement with spectral models based on the single ignition point deflagration N1def model of \cite{Fink2014}. We find several new pieces of evidence to support a weak/failed deflagration explosion model for SN~2024pxl:
\begin{itemize}
    \item The MIR forbidden emission line morphologies of both IGEs and IMEs are centrally peaked, with FWHM clustered around 8000\kms, consistent with ejecta that are well mixed on large scales through turbulent convection. 
    \item The isolated \ion{O}{1}~2.76\um\ line has a broad emission base, topped by a narrow blueshifted component. We interpret this morphology as corresponding to the channel and inner region of unburned material predicted for one-sided deflagrations from low numbers of ignition points from \cite{Fink2014}.
    \item We detect narrow, centrally peaked [\ion{Ne}{2}] at 12.81\um. Ne is nucleosynthesized at low densities and must be mixed into the center. Violent WD-WD merger models predict a significantly wider Ne line, and with the added context of other observables (e.g., light-curve properties), we find this Ne line supports a pure deflagration.
    \item Our radiative transfer model spectrum of the single-ignition-point deflagration model, N1def, from \cite{Fink2014} predicts the lines that we observe in SN~2024pxl. Artificial ``correction'' of the model continuum with a modified blackbody hints that additional flux from a central remnant may be needed.
\end{itemize}

The origin of SN~2024vjm remains more ambiguous; however, a weak deflagration explosion, possibly of a hybrid C/O/Ne WD appears promising. As in SN~2024pxl, we observe centrally peaked emission from LMEs, IMEs, and IGEs, which suggests that the ejecta are mixed. However, the luminosity and ejecta velocities in SN~2024vjm are too low for the deflagration models of \cite{Fink2014} and \cite{Lach2022}. For example, the weakest deflagration model from \cite{Fink2014} with only one ignition point, N1def, creates 0.084~$M_\odot$ of ejecta and 0.035~$M_\odot$ of $^{56}$Ni, far larger than the $M_\mathrm{Ni}\approx8\times10^{-4}~M_\odot$ measured from the light curve of the similarly low-luminosity SN~2021fcg \citep{Karambelkar2021}. 

\cite{Kromer2015} find that lower masses of $^{56}$Ni and total ejecta can be produced by the deflagration of a hybrid C/O/Ne WD (N5def\_hybrid model). The MIR [\ion{Ne}{2}]~12.81\um\ line might provide a way to compare the amount of Ne present in SN~2024pxl and SN~2024vjm and test this theory where the additional mass in the C/O/Ne WD quenches the deflagration, resulting in lower luminosity. This comparison will be better done with late-time MIR data (JWST-GO-6580, PI~S.~W.~Jha; JWST-DD-9321, PI~E.~Baron) \citep{Jha2024, Baron2025} when the [\ion{Ne}{2}] is fully nebular and SN~2024pxl and SN~2024vjm are in more similar stages of their evolution. The N5def\_hybrid simulation creates a transient with $M_V\approx-14.2$~mag, similar in luminosity to faint transients such as SN~2008ha and SN~2010ae; however, the total ejecta mass produced is too small for these objects by roughly a factor of 10 (0.014~$M_\odot$) and so the simulated transient evolves too quickly. 

As noted by \cite{Karambelkar2021}, fewer ignition points could help reduce the production of $^{56}$Ni, but the resulting transient would evolve even more rapidly. \cite{Karambelkar2021} therefore favor a merger of the kind described by \cite{Kashyap2018} to explain SN~2021fcg; however, their argument leans on photometric behavior. \cite{Ritter2021} and \cite{Lykou2023} also favor a double-degenerate merger scenario for SN~Iax based on analysis of the historical SN~1181 and its likely remnant Pa~30. It is unknown if such mergers can produce spectra similar to our observations, and future modeling should investigate this possibility. 

Current models appear inconsistent with the observed properties of extremely low-luminosity SN~Iax, but we encourage future modeling efforts that combine the deflagration mechanism and the effects of a remnant to attempt to produce transients with luminosity and duration akin to SN~2024vjm. In particular, reevaluation of the N5def\_hybrid model \citep{Kromer2015} with inclusion of remnant effects may be a promising avenue. Owing to the spectroscopic similarity to SN~2024pxl as well as to our radiative transfer N5def\_hybrid model, across optical$+$NIR$+$MIR wavelengths, we favor a weak/failed deflagration leaving behind a remnant as the origin of SN~2024vjm, but we acknowledge that current deflagration models do not explain the lowest luminosity objects well.

The panchromatic dataset presented here, accompanied by thorough ground-based photometric and spectroscopic follow-up observations of SN~2024pxl presented by \cite{Singh2025} and \cite{Hoogendam2025}, is unparalleled for SN~Iax to-date. We encourage the use of this dataset in additional analyses and for comparison to future modeling including the effects of a remnant.

These panchromatic spectra are also novel as multilayer probes of the ejecta structure, displaying photospheric P-Cygni profiles from outer layers in the optical and NIR as well as forbidden emission from interior layers in the MIR. Further detailed spectral modeling of the data might be able to stringently constrain densities, temperatures, ionization states, and composition of the ejecta, as well as distinguish between radioactive and stable Ni. Our observations highlight the power of \textit{JWST} combined with ground-based facilities to advance our understanding of the astrophysical origins of SN.

% \begin{acknowledgements}

\vspace{1.5cm}

This work is based on observations made with the NASA/ESA/CSA \textit{JWST} as part of programs \#05232, \#06580, and \#06811. We thank Shelly Meyett and Milo Docher for their consistently excellent work scheduling the \textit{JWST} observations, Sarah Kendrew, Ian Wong, and Katherine Murray for assistance with the MIRI observations, and Glenn Wahlgren, Nimisha Kumari, and Norbert Pirzkal for help with the NIRSpec observations. 

The data were obtained from the Mikulski Archive for Space Telescopes at the Space Telescope Science Institute (STScI), which is operated by the Association of Universities for Research in Astronomy (AURA), Inc., under National Aeronautics and Space Administration (NASA) contract NAS 5-03127 for \textit{JWST}. Support for this program at Rutgers University was provided by NASA through grants JWST-GO-05232.001, JWST-GO-06580.001, and JWST-GO-06811.001.

We thank the anonymous referee for insightful comments which improved the manuscript. We thank D.~John~Hillier for useful discussions regarding the atomic data used in CMFGEN, and Stuart~Sim for helpful comments on the manuscript.

L.A.K. is supported by a CIERA Postdoctoral Fellowship. M.S. acknowledges financial support provided under the National Post Doctoral Fellowship (N-PDF; File Number PDF/2023/002244) by the Science \& Engineering Research Board (SERB), Anusandhan National Research Foundation (ANRF), Government of India.

The SALT observations were obtained with Rutgers University program 2024-1-MLT-004 (PI L.~A.~Kwok). We are grateful to SALT Astronomer Rosalind Skelton for taking these data. Data reported here were obtained in part at the MMT Observatory, a joint facility of the University of Arizona and the Smithsonian Institution. We sincerely thank the MMT observers and staff for their accommodation of our strict timing requirements to ensure they coincided with {\it JWST} observations. We are grateful to the staff of IAO, Hanle, CREST, and Hosakote, who made these observations possible. The facilities at IAO and CREST are operated by the Indian Institute of Astrophysics, Bangalore. We thank the Subaru staff for the data taken by the Subaru Telescope (S23A-023).

Observations from coauthor A.J.M. were made under the aegis of the ASTRAL (Astronomy/STEM Alliance with Lick Observatory) consortium, supported by a generous grant from the Gordon and Betty Moore Foundation (PI B. Macintosh).

Some of the data presented herein were obtained at the W.~M. Keck Observatory, which is operated as a scientific partnership among the California Institute of Technology, the University of California, and NASA. The Observatory was made possible by the generous financial support of the W.~M. Keck Foundation. The authors wish to recognize and acknowledge the very significant cultural role and reverence that the summit of Maunakea has always had within the indigenous Hawaiian community. We are most fortunate to have the opportunity to conduct observations from this mountain.

A major upgrade of the Kast spectrograph on the Shane 3~m telescope at Lick Observatory, led by Brad Holden, was made possible through generous gifts from the Heising-Simons Foundation, William and Marina Kast, and the University of California Observatories. Research at Lick Observatory is partially supported by a generous gift from Google.

Based (in part) on data acquired at the ANU 2.3~m telescope. The automation of the telescope was made possible through an initial grant provided by the Centre of Gravitational Astrophysics and the Research School of Astronomy and Astrophysics at the Australian National University and through a grant provided by the Australian Research Council through LE230100063. We acknowledge the traditional custodians of the land on which the telescope stands, the Gamilaraay people, and pay our respects to elders past and present.

This work is based in part on observations collected at the European Southern Observatory under ESO programme 114.27JL.001. Partly based on observations made with the Nordic Optical Telescope, owned in collaboration by the University of Turku and Aarhus University. This work also makes use of data gathered with the 6.5~m Magellan telescopes at Las Campanas Observatory, Chile. A FIRE spectrum was obtained through A.P.'s prior support by a Carnegie Fellowship. This work makes use of data from the Las Cumbres Observatory global network of telescopes. The LCO group is supported by NSF grants AST-1911151 and AST-1911225.

A.A.M. and N.R. are supported by DoE award \#DE-SC0025599. MMT and W. M. Keck Observatory access for N.R. and C.L. was supported by Northwestern University and the Center for Interdisciplinary Exploration and Research in Astrophysics (CIERA).
A.V.F.’s group at UC Berkeley received financial assistance from the Christopher R. Redlich Fund, as well as donations from Gary and Cynthia Bengier, Clark and Sharon Winslow, Alan Eustace, William Draper, Timothy and Melissa Draper, Briggs and Kathleen Wood, Sanford Robertson (W.Z. is a Bengier-Winslow-Eustace Specialist in Astronomy, T.G.B. is a Draper-Wood-Robertson Specialist in Astronomy, Y.Y. was a Bengier-Winslow-Robertson Fellow in Astronomy), and numerous other donors.

K. Maguire acknowledges funding from Horizon Europe ERC grant 101125877. J.H.T. acknowledges support from EU H2020 ERC grant 758638. T.T. acknowledges support from NSF grant AST-2205314 and the NASA ADAP award 80NSSC23K1130. K. Maede acknowledges support from JSPS KAKENHI grants JP24KK0070, JP24H01810, and JP20H00174, and from JSPS Bilateral Joint Research Project (JPJSBP120229923). J.V. is supported by NKFIH-OTKA grant K142534. G.C.A. thanks the Indian National Science Academy for support under the INSA Senior Scientist Programme. M.R.S. is supported by the STScI Postdoctoral Fellowship. 

K. Mistra acknowledges support from the BRICS grant DST/ICD/BRICS/Call-5/CoNMuTraMO/2023 (G) funded by the DST, India. R.D. acknowledges funds by ANID grant FONDECYT Postdoctorado \#3220449. B.B. received support from the Hungarian National Research, Development and Innovation Office grants OTKA PD-147091. L.G. acknowledges financial support from AGAUR, CSIC, MCIN, and AEI 10.13039/501100011033 under projects PID2023-151307NB-I00, PIE 20215AT016, CEX2020-001058-M, ILINK23001, COOPB2304, and 2021-SGR-01270. H.K. was funded by the Research Council of Finland projects 324504, 328898, and 353019. J.A.V. acknowledges the Postgraduate School of the Universidad de Antofagasta for its support and allocated grants. C.L. acknowledges support from DOE award DE-SC0010008 and NSF award AST-2407567 to Rutgers University.

J.E.A. is supported by the international Gemini Observatory, a program of NSF's NOIRLab, which is managed by the Association of Universities for Research in Astronomy (AURA) under a cooperative agreement with the NSF, on behalf of the Gemini partnership of Argentina, Brazil, Canada, Chile, the Republic of Korea, and the United States of America.

A.F. acknowledges support by the European Research Council (ERC) under the European Union’s Horizon 2020 research and innovation program (ERC Advanced Grant KILONOVA \#885281) and the State of Hesse within the Cluster Project ELEMENTS. Time-domain research by the University of Arizona team and D.J.S. is supported by NSF grants 2108032, 2308181, 2407566, and 2432036 and the Heising-Simons Foundation under grant \#2020-1864. K.A.B. is supported by an LSST-DA Catalyst Fellowship; this publication was thus made possible through the support of grant 62192 from the John Templeton Foundation to LSST-DA. N.F. acknowledges support from the NSF Graduate Research Fellowship Program under grant DGE-2137419. A.C.G. and the Fong Group at Northwestern acknowledges support by the NSF under grants AST-1909358, AST-2206494, AST-2308182, and CAREER grant AST-2047919. This work was supported by the ``Action Thématique de Physique Stellaire'' (ATPS) of CNRS/INSU PN Astro co-funded by CEA and CNES. Time-domain research by the University of California, Davis team and S.V. is supported by NSF grant AST-2407565.

This work made use of the Heidelberg Supernova Model Archive (HESMA), https://hesma.h-its.org.

\facilities{AAVSO, ANU (WiFeS), ASAS-SN, ATLAS, GTC (EMIR), JWST (NIRSpec/MIRI), Keck:I (LRIS), Keck:II (NIRES), Keck:II (DEIMOS), LCO/GSP, Magellan (IMACS), MMT (Binospec), SALT (RSS), Shane (Kast), SOAR (Goodman), Subaru (FOCAS), UH (SNIFS), VLT (XSHOOTER), ZTF}

\software{Astropy \citep{astropycollaboration_astropy:_2013, astropycollaboration_astropy_2018, AstropyCollaboration2022}, 
Matplotlib \citep{hunter_matplotlib:_2007}, 
NumPy \citep{oliphant_guide_2006}, PyRAF \citep{Pyraf}, PySALT \citep{PySALT}, dust extinction \citep{karl_gordon_2022_6397654}, jdaviz \citep{jdadf_developers_2022_7255461}, jwst \citep{Bushouse_JWST_Calibration_Pipeline_2022}, UltraNest \citep{Buchner2021, johannes_buchner_2022_7053560}, {\tt YSE-PZ} \citep{CoulterZenodo, CoulterYSEPZ}, CMFGEN \citep{Hillier2012}
}

\appendix

\section{MIRI/LRS Wavelength Calibration \label{sec:LRS_cal}}
There is a known issue with the MIRI/LRS wavelength calibration that has gone through several iterations of improvement. In comparing the LRS spectra of SN~2024pxl to its MRS spectrum, for which the wavelength solution is accurate, we find a clear discrepancy at wavelengths $\lambda\lesssim$8\um. Our MRS spectrum was taken only 5~days before our final LRS spectrum for SN~2024pxl, at $+$37 and $+$42~days, respectively. At this phase, the forbidden MIR lines at wavelengths $\lambda\gtrsim$8\um\ agree very closely with minimal velocity evolution between these epochs. Therefore, to correct the wavelength solution of our LRS spectra, we match the peaks of each clear SN emission line between 5--12.5\um\ between the MRS spectrum and the $+$42~day LRS spectrum. To do so, we centroid the peaks near 5.7, 6.3, 6.6, 7.0, 7.4, 8.5, 10.5, and 12.0\um\ in both the MRS and final LRS spectrum with a bootstrapping method. We then fit a linear correction to the difference in wavelength of the peaks between the LRS and MRS spectra, weighted by the uncertainties. We correct all LRS spectra by this fit, with the equation given below,

\begin{equation*}
    \Delta\lambda=0.00376~\lambda_\mathrm{LRS} - 0.0431\,\mu\mathrm{m}
\end{equation*}
where $\Delta\lambda= \lambda_\mathrm{LRS} - \lambda_\mathrm{MRS}$.

\begin{figure}[b]
    \centering
    \includegraphics[width=0.5\linewidth]{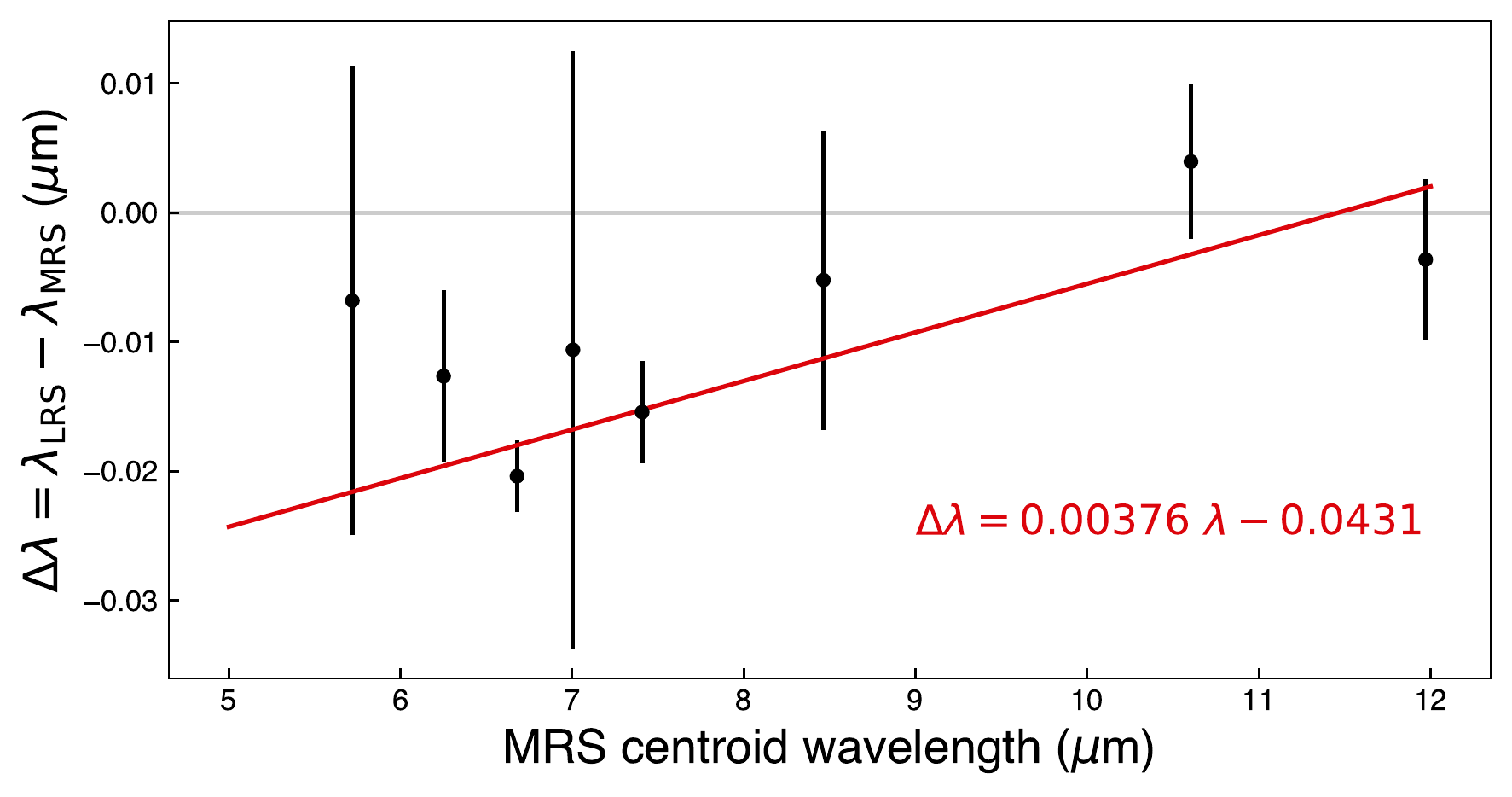}
    \caption{The discrepancy in the LRS wavelength solution introduces some additional uncertainties to the measured line velocities and widths.}
    \label{fig:wave_corr}
\end{figure}

\begin{figure}
    \centering
    \includegraphics[width=\linewidth]{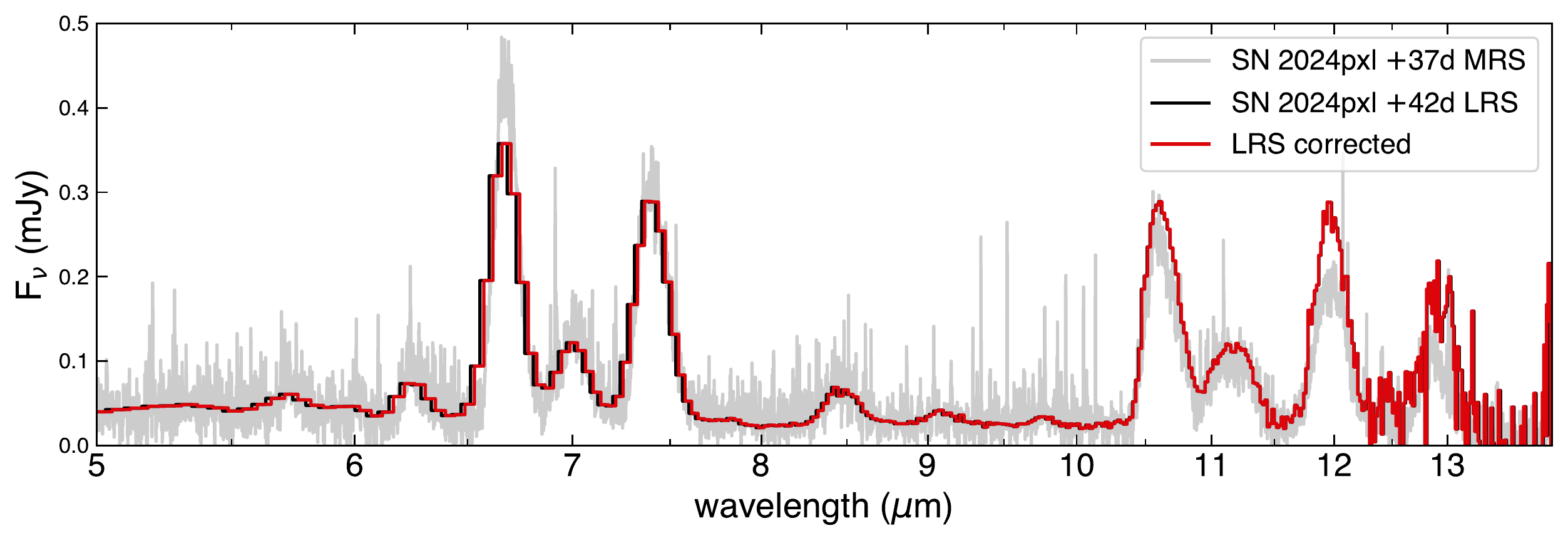}
    \caption{Comparison of the MRS spectrum of SN~2024pxl at $+$37~days (gray), the LRS spectrum at $+$42~days (black), and the LRS spectrum with our wavelength correction applied (red).}
    \label{fig:LRS_MRS_corr}
\end{figure}

We then take our wavelength correction and apply it to the earlier LRS epochs. While there is still some uncertainty in this correction due to the changing velocities of the SN ejecta over time, this is likely within the uncertainty from the LRS resolution. \autoref{fig:wave_corr} and \autoref{fig:LRS_MRS_corr} show our wavelength correction. We suggest that additional calibration between LRS and MRS should be done, potentially by observing the same thermonuclear SN with forbidden emission lines in the 5--8\um\ range contemporaneously with both LRS and MRS to accurately calibrate the short wavelengths for all future LRS observations.

\section{Line Identification \label{sec:line_id_appendix}}

The model from which we base our line identifications has been calculated at 50~days post-explosion ($\sim+37$~days post maximum-light) and includes both permitted and forbidden line transitions. The atomic data used here are the same used by \citet{Blondin2023}. By $+37$~days post-maximum in SN~2024pxl, the velocities are narrow enough that the peak of most individual contributing lines can be distinguished, and we can match them to model lines. We use the predicted strengths from the model line list to guide our identifications and check that the velocity offset of the line peak is consistent with other nearby lines and/or those seen in other lines of the same ion. 

To further improve our confidence in our line identifications and determine how much a particular ion contributes when multiple potential lines are very close in wavelength, we calculate versions of the model spectrum omitting one ion at a time. Then we divide by the full-ion model spectrum and subtract from 1 to obtain the contribution of each individual ion. We note that by removing an ion from the calculation, we remove a source of absorption and/or emission which can change the overall spectrum, not just a single ion's contribution. Our method gives a close approximation of the contribution of an individual ion, but it may not be exactly the same calculation as if the ion were included. The complete model line-list used will be made available on Zenodo\footnote{\url{https://zenodo.org/communities/snrt/}}.

For our line identifications of [\ion{Mg}{2}], we check that omitting \ion{Mg}{2} in the model spectrum calculation creates a flux deficit compared to the full model spectrum at these wavelengths (specifically 3.09, 4.76, and 9.71\um), increasing our confidence that \ion{Mg}{2} is indeed responsible for the observed emission.

\begin{table}[h]
\centering
\begin{tabular}{c c c}
\toprule
Phase (days) & v$_\mathrm{phot}$ (\kms) & FWHM (\kms) \\
\midrule
SN~2024pxl & \\
8 & $-$4550 $\pm$ 270 & 1980 $\pm$ 670 \\
10 & $-$4260 $\pm$ 140 & 1890 $\pm$ 260 \\
18 & $-$3650 $\pm$ 1110 & 2500 $\pm$ 260 \\
23 & $-$3110 $\pm$ 100 & 2600 $\pm$ 130 \\
34 & $-$2510 $\pm$ 200 & 2580 $\pm$ 270 \\
36 & $-$2390 $\pm$ 170 & 2950 $\pm$ 190 \\
37 & $-$2530 $\pm$ 250 & 2540 $\pm$ 440 \\
42 & $-$2380 $\pm$ 1340 & 3460 $\pm$ 830 \\
SN~2024vjm \\
12 & $-$1180 $\pm$ 60 & 1430 $\pm$ 160 \\
\bottomrule
\end{tabular}

\caption{Co II Line Velocities}
\label{tab:coii_vels}
\end{table}

\begin{table}[h]
\centering
\begin{adjustwidth}{-2.5cm}{}
\begin{tabular}{c|c c|c c|c c|c c}
\toprule
 & \multicolumn{2}{c}{Epoch 1 ($+$11 d)} & \multicolumn{2}{c}{Epoch 2 ($+$21 d)} & \multicolumn{2}{c}{Epoch 3 ($+$37 d)} & \multicolumn{2}{c}{Epoch 4 ($+$42 d)}\\
 Ion & FWHM (\kms) & v$_\mathrm{phot}$ (\kms) & FWHM (\kms) & v$_\mathrm{phot}$ (\kms) & FWHM (\kms) & v$_\mathrm{phot}$ (\kms) & FWHM (\kms) & v$_\mathrm{phot}$ (\kms) \\
\midrule
{[Co II]} & 8800 $\pm$ 500 & 600 $\pm$ 200 & 8600 $\pm$ 200 & 900 $\pm$ 100 & 6880 $\pm$ 50 & 360 $\pm$ 20 & 7500 $\pm$ 200 & 700 $\pm$ 100 \\
{[Ni II]} & 9300 $\pm$ 700 & 200 $\pm$ 300 & 8400 $\pm$ 300 & 400 $\pm$ 100 & 6030 $\pm$ 20 & 290 $\pm$ 10 & 7900 $\pm$ 200 & 200 $\pm$ 100 \\
{[Ar II]} & 8500 $\pm$ 600 & $-$1300 $\pm$ 200 & 8400 $\pm$ 600 & $-$800 $\pm$ 300 & 9450 $\pm$ 120 & $-$170 $\pm$ 50 & 7400 $\pm$ 1000 & $-$200 $\pm$ 400 \\
{[Ar III]} & 9300 $\pm$ 1300 & 3200 $\pm$ 600 & 8000 $\pm$ 2900 & 2500 $\pm$ 1400 & --- & 1550 $\pm$ 420 & 8300 $\pm$ 3400 & 2200 $\pm$ 1600 \\
{[Ni III]} & 10200 $\pm$ 200 & 1500 $\pm$ 100 & 9100 $\pm$ 200 & 1300 $\pm$ 100 & 7860 $\pm$ 40 & 780 $\pm$ 10 & 8300 $\pm$ 300 & 800 $\pm$ 100 \\
{[Co III]} & 8400 $\pm$ 400 & 1400 $\pm$ 200 & 7600 $\pm$ 200 & 1100 $\pm$ 100 & 8180 $\pm$ 70 & 680 $\pm$ 30 & 8000 $\pm$ 200 & 400 $\pm$ 100 \\
{[Ni IV]} & --- & 2000 $\pm$ 500 & --- & 600 $\pm$ 800 & --- & $-$480 $\pm$ 120 & 8500 $\pm$ 2200 & 1000 $\pm$ 700 \\
{[Mg II]} & 6800 $\pm$ 300 & 100 $\pm$ 100 & 6000 $\pm$ 1000 & 600 $\pm$ 300 & 12000 $\pm$ 0 & 630 $\pm$ 280 & 4700 $\pm$ 4000 & $-$100 $\pm$ 1400 \\
{[Ne II]} & --- & --- & --- & --- & 9260 $\pm$ 440 & 690 $\pm$ 80 & 7500 $\pm$ 300 & 1000 $\pm$ 100 \\
{[Co IV]} & --- & --- & --- & --- & 6210 $\pm$ 300 & $-$1130 $\pm$ 170 & --- & --- \\
\bottomrule
\end{tabular}

\caption{SN 2024pxl Forbidden Line Velocities}
\label{tab:24pxl_neb_vels}
\end{adjustwidth}
\end{table}

\begin{table}[h]
\centering
\begin{tabular}{c c c}
\toprule
 Line & FWHM (\kms) & v$_\mathrm{phot}$ (\kms) \\
\midrule
{[Mg II]} 4.76 $\mu$m & 2700 $\pm$ 200 & 20 $\pm$ 60 \\
{[Ni II]} 6.64 $\mu$m & 3300 $\pm$ 100 & 380 $\pm$ 20 \\
{[Ar II]} 6.98 $\mu$m & 4200 $\pm$ 200 & 220 $\pm$ 30 \\
{[Ni III]} 7.35 $\mu$m & 3400 $\pm$ 400 & $-$80 $\pm$ 150 \\
{[Mg II]}? 7.45 $\mu$m & 5200 $\pm$ 300 & 450 $\pm$ 100 \\
{[Co II]} 10.52 $\mu$m & 3100 $\pm$ 300 & 400 $\pm$ 70 \\
{[Ni II]} 10.68 $\mu$m & 3700 $\pm$ 800 & 130 $\pm$ 550 \\
{[Co III]} 11.89 $\mu$m & 5800 $\pm$ 300 & 350 $\pm$ 90 \\
{[Mg II]}? 12.3 $\mu$m & 4300 $\pm$ 300 & $-$520 $\pm$ 80 \\
{[Ne II]} 12.81 $\mu$m & 4300 $\pm$ 200 & $-$180 $\pm$ 60 \\
{[Co II]} 14.74, 15.46 $\mu$m & 3000 $\pm$ 100 & 560 $\pm$ 20 \\
\bottomrule
\end{tabular}

\caption{SN 2024vjm Forbidden Line Velocities}
\label{tab:24vjm_neb_vels}
\end{table}

\begin{table}[h]
\centering
\begin{tabular}{c c c c c}
 & & Peak F$_\nu$ (mJy) & & \\
\midrule
Line & Epoch 1 ($+$11 d) & Epoch 2 ($+$21 d) & Epoch 3 ($+$37 d) & Epoch 4 ($+$42 d) \\
\midrule
{[Ni II]} 6.64 $\mu$m & 0.064 $\pm$ 0.002 & 0.175 $\pm$ 0.003 & 0.406 $\pm$ 0.005 & 0.326 $\pm$ 0.006 \\
{[Ar II]} 6.98 $\mu$m & 0.076 $\pm$ 0.003 & 0.079 $\pm$ 0.003 & 0.118 $\pm$ 0.006 & 0.092 $\pm$ 0.006 \\
{[Ni III]} 7.35 $\mu$m & 0.126 $\pm$ 0.002 & 0.174 $\pm$ 0.003 & 0.289 $\pm$ 0.005 & 0.261 $\pm$ 0.006 \\
{[Ni IV]} 8.05 $\mu$m & 0.024 $\pm$ 0.002 & 0.031 $\pm$ 0.004 & 0.058 $\pm$ 0.015 & 0.044 $\pm$ 0.006 \\
{[Ar III]} 8.98 $\mu$m & 0.016 $\pm$ 0.002 & 0.015 $\pm$ 0.003 & --- & 0.021 $\pm$ 0.004 \\
{[Mg II]} 9.71 $\mu$m & 0.056 $\pm$ 0.002 & 0.04 $\pm$ 0.004 & --- & 0.017 $\pm$ 0.004 \\
{[Co II]} 10.5 $\mu$m & 0.045 $\pm$ 0.002 & 0.129 $\pm$ 0.003 & 0.238 $\pm$ 0.005 & 0.271 $\pm$ 0.006 \\
{[Ni III]} 11.0 $\mu$m & 0.037 $\pm$ 0.002 & 0.064 $\pm$ 0.002 & 0.104 $\pm$ 0.008 & 0.104 $\pm$ 0.004 \\
{[Co III]} 12.0 $\mu$m & 0.054 $\pm$ 0.009 & 0.119 $\pm$ 0.003 & 0.179 $\pm$ 0.011 & 0.274 $\pm$ 0.006 \\
\bottomrule
\end{tabular}

\caption{SN~2024pxl Forbidden Line Peak Flux}
\label{tab:peak_flux}
\end{table}

% \begin{figure*}
%     \centering
%     \includegraphics[width=\linewidth]{blackbody_fits.png}
%     \caption{Caption}
%     \label{fig:BB_temps}
% \end{figure*}

\clearpage

\normalsize
\bibliography{zotero_abbrev}

\end{document}